\documentclass[]{interact}

\usepackage{epstopdf}% To incorporate .eps illustrations using PDFLaTeX, etc.
\usepackage[textwidth=1.8cm, textsize=tiny]{todonotes}
\usepackage{makecell}
\usepackage{cmll}
\usepackage{graphicx}
\usepackage{booktabs}
\usepackage{multicol,multirow}

\usepackage{booktabs}
\usepackage{siunitx}
\usepackage{bbm}
\usepackage{url}
\makeatletter
\g@addto@macro\UrlBreaks{\do\-\do\/\do\_\do\.\do\?\do\&\do\=\do\#}
\makeatother
\usepackage[table]{xcolor} % loads colortbl
\definecolor{lightgrayrow}{gray}{0.92}

\usepackage[pdftex]{changebar}
\usepackage[longnamesfirst,sort]{natbib}% Citation support using natbib.sty
\bibpunct[, ]{(}{)}{;}{a}{,}{,}% Citation support using natbib.sty
\renewcommand\bibfont{\fontsize{10}{12}\selectfont}% To set the list of references in 10 point font using natbib.sty

\theoremstyle{plain}% Theorem-like structures provided by amsthm.sty

\theoremstyle{definition}

\theoremstyle{remark}

\usepackage{subcaption}

\begin{document}

\articletype{Research article}% Specify the article type or omit as appropriate

% \title{Spatiotemporal data integration framework for improved flood cost and occurrence modelling in house insurance}

\title{Bayesian spatial modelling framework for assessing residential flood risk in property insurance}
% performance and mutualisation implications
\author{
\name{Mulah Moriah \textsuperscript{a,b}\thanks{Corresponding author:  Mulah Moriah. Email: mulah.moriah@etudiant.univ-brest.fr}, Franck Vermet\textsuperscript{a,b}, Pierre Ailliot\textsuperscript{a,b}, Philippe Naveau\textsuperscript{c} and Juliette Legrand\textsuperscript{a,b}}
\affil{\textsuperscript{a}Univ Brest, CNRS UMR 6205, Laboratoire de Mathématiques de Bretagne Atlantique, France; \textsuperscript{b}Euro-Institut d'Actuariat Jean Dieudonné (EURIA), UBO, Brest, France; \textsuperscript{c}Laboratoire des sciences du climat et de l'environnement (EstimR),
Université Paris-Saclay, CNRS, CEA, UVSQ, 91191 Gif-sur-Yvette, France}
}

\maketitle

\begin{abstract}
Spatial heterogeneity in insurance risk modelling is often represented using coarse areal structures and covariates, which can obscure fine-scale patterns critical for accurate risk assessment. This study introduces a point-referenced Bayesian framework to model the occurrence and severity of claims at the policyholder level, avoiding reliance on predefined geographic aggregation. Drawing on a large French insurance portfolio combined with high-resolution environmental variables, rainfall records, and institutional hazard maps, we compare a benchmark Generalised Linear Model with several discrete Bayesian specifications, including independent random effects, intrinsic conditional autoregressive (iCAR) and Besag-York-Mollié (BYM) models, and a continuously indexed Gaussian random field constructed using the Stochastic Partial Differential Equation (SPDE) approach.

\noindent Inference is performed using the Integrated Nested Laplace Approximation (INLA), enabling efficient estimation of latent spatial fields and non-linear covariate effects. Our results show that accounting for spatial dependence substantially improves occurrence modelling, while gains in severity prediction are more limited. The SPDE formulation further outperforms areal models by capturing sub-municipal risk gradients and reducing artefacts induced by arbitrary geographic partitioning. By conditioning on detailed building-level attributes, we isolate the contribution of latent spatial effects, refine the interpretation of observed covariates, and improve the allocation of risk premiums across the portfolio. In addition to improved predictive performance, the framework provides coherent uncertainty quantification and supports tail-risk assessment. To our knowledge, this is the first application of point-referenced INLA-SPDE models to flood insurance, offering a scalable statistical alternative for pricing and managing risks with strong spatial structure.

\end{abstract}

\begin{keywords}
House insurance flood risk; INLA-SPDE; Spatial modelling; Pricing
\end{keywords}

\section{Introduction}

\subsection{Context and challenges in flood risk pricing}

Property and casualty ratemaking seeks to determine premiums that match the anticipated cost of future claims for a specific risk. In practical applications, insurers commonly use Generalised Linear Models (GLMs) because they offer a clear and established compromise between interpretability, regulatory approval, and predictive performance \citep{Boa06,Gol16}. In this setting, claim frequency and claim severity are usually modelled separately and then combined to derive the expected pure premium, drawing on information about building characteristics (e.g. size, construction year, amenities, insured amounts), policyholder features and, increasingly, environmental and geospatial data.

Spatial effects have long been recognized as fundamental determinants of insurance risk \citep{Tay89,Bos94}, especially for hazards that are inherently spatial, such as floods. Flood risk is inherently tied to location and results from complex interactions between hazard, exposure, and vulnerability. Prior studies have repeatedly highlighted the value of incorporating external data sources to capture these distinct components of flood risk. Such data commonly comprise meteorological and hydrological measures \citep{Mer13,Spe14,Spe11,Tan20,Ber19}, geographic and environmental attributes \citep{Tor17,Mob21,Cha21,Dri23,Cep22}, detailed building characteristics \citep{Mer04,And13,Mar21,Gra14}, and socio-economic variables serving as proxies for vulnerability \citep{Ass14}.

From the insurer's perspective, claims are recorded at the building or contract level, and numerous studies indicate that high-resolution information is crucial to prevent aggregation bias. Using coarse spatial units can underestimate risk in highly exposed micro-locations and overstate it in others, resulting in inefficient pricing and increased adverse selection \citep{Bou22,Mor25}. These issues are intensified by the documented increase in flood-related losses, jointly driven by climate change and socio-economic growth \citep{Ass21}. %In France, flood coverage is embedded in the Catastrophes Naturelles (CatNat) regime, a compulsory public--private scheme funded by a uniform surcharge on property insurance premiums and backstopped by the state reinsurer CCR. Because the surcharge is flat, building-level risk assessment remains essential for underwriting, portfolio management, and internal capital allocation.

In operational settings, geographic risk is typically introduced through discrete territorial segmentations such as administrative areas or actuarial zones derived from GLMs, credibility models, smoothing procedures or clustering algorithms \citep{Jen08,Ema11,Riv21}. A common workflow is to first calibrate models without geographic variables and then attribute the remaining unexplained variation to an unobserved geographic component. The residuals can then be estimated using the available geographic data and then smoothed and/or grouped into clusters \citep{Ema11}. Although this strategy helps interpretability and meets regulatory standards, it makes only partial use of the growing richness of spatial data. In particular, it has difficulty representing non-linear effects and often forces continuous variables to be discretised \citep{Den04}. In addition, spatial terms are introduced as additive adjustments rather than being fully integrated, which implicitly assumes that spatial information is independent of other predictors. Furthermore, as noted by \citet{Gol16}, GLMs place full credibility in training data regardless of segment size and posit that, conditional on observed covariates, the outcomes are not correlated. This is clearly violated when unobserved latent factors influence clusters of nearby risks, as is common for climate-related perils. These shortcomings point to the need for modelling approaches that jointly capture covariate effects and spatial dependence within a unified estimation framework.

\subsection{Extending spatial flood risk modelling with INLA-SPDE}
To address these challenges, the actuarial literature has increasingly shifted towards fully integrated modelling frameworks, in which claim frequency and severity are modelled separately while spatial dependence is incorporated directly into each component \citep{Dim02,Den04,Gsc07,Ass14,Tuf19}. Earlier multi-stage or ratio-based approaches \citep{Tay89,Bos94,Smy02,Jor94} have been largely left out, since modelling frequency and severity individually improves interpretability and permits distinct covariate and spatial structures.

An important methodological advance was the transition from GLMs to Bayesian generalised linear mixed models (GLMMs), where spatial effects are introduced as latent random components \citep{Dim02}. This framework relaxes conditional independence and accommodates unobserved heterogeneity \citep{Gol16}. In a similar vein, \citet{Den04} proposed generalised additive mixed models to flexibly represent non-linear covariate effects within a unified estimation scheme. Spatial dependence in these models is typically encoded via areal structures such as the Besag-York-Mollié (BYM) or ICAR specifications \citep{Bes91}, which induce smoothing across adjacent regions. However, these methods depend on fixed geographic units, limiting spatial granularity and potentially conflicting with the inherently point-level nature of insurance claims data.

This paper develops a fully integrated Bayesian framework that moves away from areal representations of space. We model claim frequency and severity separately at the individual policyholder level, avoiding any aggregation into administrative or actuarial regions, and we draw on both insurance records and several high-resolution external datasets. This point-referenced perspective, in line with recent work \citep{Riv25,Wah22}, naturally matches individual risk pricing and allows full use of detailed insurance and environmental information. This configuration provides an appropriate setting for evaluating the marginal contribution of latent spatial processes. By conditioning on high-resolution building and environmental characteristics, features often omitted or aggregated in previous studies, we can empirically determine explicit spatial effects contributions.

We estimate the model using the Integrated Nested Laplace Approximation (INLA; \citealp{Rue09,Ill12}), which provides fast and accurate inference for latent Gaussian models by leveraging the sparse structure of Gaussian Markov random fields (GMRFs). To capture spatial dependence continuously over the domain, we combine INLA with the Stochastic Partial Differential Equation (SPDE) approach \citep{Lin11}, which connects Gaussian random fields with Matérn covariance functions to GMRFs defined on a finite element mesh. In doing so, we extend the previous actuarial uses of INLA that have been mainly based on areal spatial models \citep{Wah22,Tuf19}. Using the INLA-SPDE framework, this study constructs a continuous representation of spatial risk, allowing inference at any desired resolution and capturing sub-municipal variability that traditional approaches tend to obscure. The framework supports joint estimation of structured latent spatial fields and complex, non-linear covariate effects within a unified Bayesian setting. To our knowledge, this is the first use of point-referenced SPDE models to investigate flood-related housing losses, demonstrating how bypassing administrative boundaries mitigates artefacts of geographic partitioning and enhances the precision of statistical climate risk assessment.

% \subsection{Agenda}
\subsection{Objective and outline of the paper}

This study aims to assess how advanced spatial modelling frameworks can enhance the statistical assessment of flood risk for property and casualty insurance. The analysis relies on a large portfolio of property insurance for mainland France over 2014-2022, integrating underwriting records with detailed georeferenced building characteristics, environmental variables, rainfall metrics and official hazard maps. Previous work on the same portfolio by \citep{Mor25} has shown that constructing and integrating such high-resolution external data into a standard GLM framework markedly improves the prediction and segmentation of both claim frequency and severity, outperforming traditional methods based on coarse risk zoning. Section~\ref{sec:data} describes the insurance portfolio and the geolocated external covariates, while a more comprehensive account of the data sources and preprocessing can be found in \citep{Mor25}.

In this paper, we advance this line of work by explicitly accounting for spatial dependence within a unified Bayesian framework. Using the same portfolio and set of covariates, we compare multiple model formulations that differ in how they represent spatial structure: a standard GLM as a benchmark, a Bayesian GAM without spatial components, models with discretely indexed spatial random effects (IID, iCAR, BYM) at the municipality level, and a continuously indexed spatial model relying on a Matérn Gaussian random field implemented via the SPDE approach. We focus in particular on the balance between predictive accuracy, model complexity, and computational cost. Section~\ref{sec:models} outlines the modelling frameworks, covering discrete and continuous spatial effects, prior choices, mesh construction, and the validation scheme.

\noindent Section~\ref{sec:results} provides an extensive comparison of the outcomes for the occurrence, cost, and pure premium, combining statistical performance metrics, spatial interpretability, portfolio-level pricing diagnostics, and uncertainty quantification. The paper ends with a discussion of the limitations of INLA-SPDE, as well as the advantages and drawbacks of employing a fine-grained modelling approach.

\begin{figure}[ht]
\centering
\includegraphics[width=0.80\linewidth]{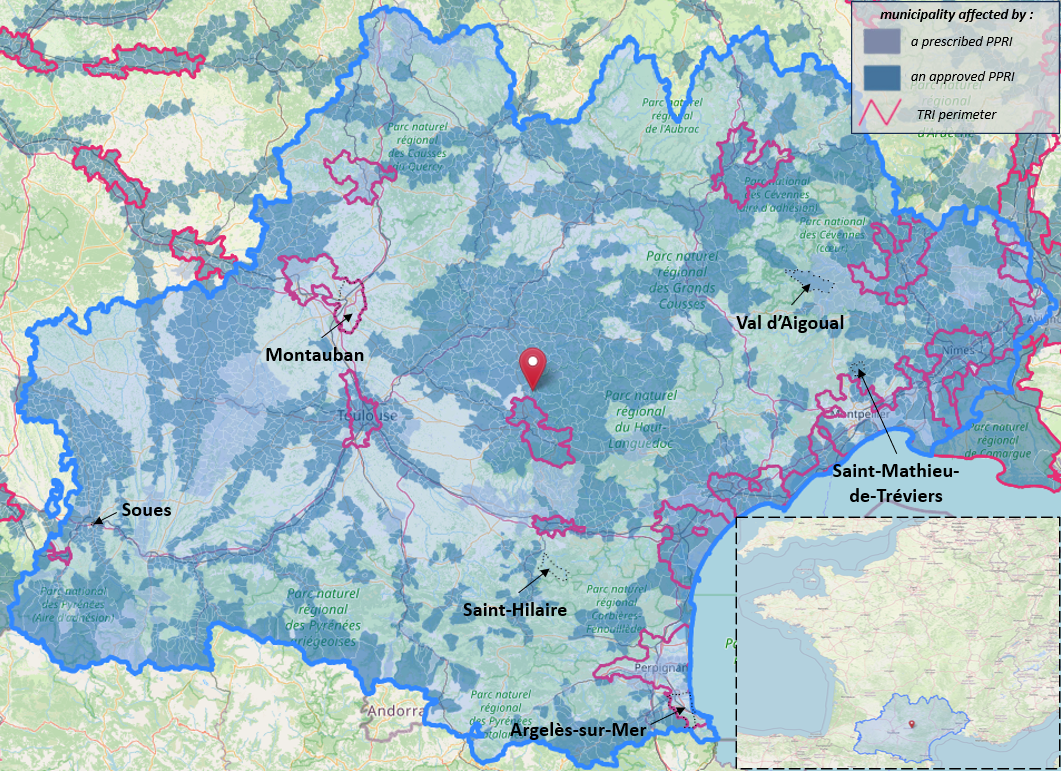}
\caption{Geographical and regulatory overview of the Occitania region. The main map shows how flood risk prevention plans (PPRI) are distributed across municipalities, along with the boundaries of areas with significant flood risk (TRI). Labelled cities indicate the key locations used in Section \ref{sec:results} analysis. The inset map (bottom right) situates the study area within France national context.}
\label{fig:occitania_desc}
\end{figure}

Note that the maps are shown only for the Occitanie region, whereas all numerical results refer to the entire portfolio. The local climate of Occitanie is shaped by Mediterranean influences and complex topography, which together produce episodes of very intense rainfall and rapidly developing flash floods. The region is also frequently affected by river flooding and is the region of France where floods generate the highest financial losses \citep{Insee24}. Figure \ref{fig:occitania_desc} locates Occitanie within mainland France and offers a spatial overview of the institutional flood risk management framework relevant to our analysis. Several municipalities are explicitly labelled, as they are used as case studies in the results section~\ref{sec:results}.

\section{Data}\label{sec:data}

The portfolio used in this study includes about 968,000 different insured properties, corresponding to 4.9 million policy records. During the observation window, roughly 10,800 flood-related claims were filed, underscoring the rarity of flood risk.
The claims considered in this study arise from pluvial flooding (direct rainfall exceeding local drainage capacity) and fluvial flooding (river overflow), which are the two dominant flood mechanisms affecting residential properties in mainland France. Flood claims are concentrated in space and time. The grouping of claims within a five-day window and between adjacent municipalities reduces the 10,800 individual claims to approximately 3,600 distinct flooding episodes. The ten largest episodes account for roughly one third of all claims, while the remainder result from numerous smaller, geographically scattered events.

For each insured property, the underwriting database reports contract and property level variables typically used in home insurance pricing. These cover characteristics such as the number of rooms and floors, declared values of content and high-value items, the presence of specific amenities, and the existence and size of outbuildings. Each building was georeferenced, making it possible to link the portfolio with external spatial datasets describing flood exposure and vulnerability.
Building-level characteristics are supplemented with fine-grained descriptors of the immediate environment. These comprise topographic and land use indicators such as slope, curvature, relative elevation compared to the surrounding terrain, soil type and permeability, density of the building within specified buffers, and distance/relative elevation to the closest watercourse. Together, these variables are designed to represent local environmental conditions that control the propagation of floods and associated damage processes.

The dataset was also enhanced with institutional hazard and regulatory maps produced by French public agencies, including the Territoires à Risque Important d'Inondation (TRI), the Plans de Prévention des Risques Inondation (PPRI), hydrographic regions, climatic zones, and the national inventory of natural disasters under the French framework.This framework operates within the Catastrophes Naturelles (CatNat) system, a mandatory public-private scheme in which flood insurance is automatically included in standard homeowners’ policies, financed through a flat surcharge on property premiums and ultimately guaranteed by the state reinsurer CCR. TRI maps are based on detailed hydraulic and topographic modelling and provide a high-quality benchmark for assessing flood hazards \citep{Cep22}, but cover only about 5\% of the portfolio due to their high production cost. PPRI maps extend to a larger fraction of the territory (roughly 30\%), but exhibit greater heterogeneity in both resolution and methodology, ranging from simplified modelling approaches to expert judgement, and most often result in large-scale hazard maps \citep{PPRI2023}.

Meteorological covariates are derived from the ERA5-Land reanalysis, which offers high-resolution rainfall time series \citep{ECMWF19}. From these data, \cite{Mor25} constructed indicators that describe both precipitation intensity and accumulation, aggregated annually and within windows centred on the observed claim dates. These variables provide a dynamic view of the meteorological drivers of flooding, complementing static information on hazard, exposure, and vulnerability. To represent the key meteorological drivers of floods, the authors defined the Most Intense Local Rainfall Event (MILRE) and its annual aggregate (ann\_MILRE). As detailed in Appendix \ref{app:rainfall}, they did not use raw rainfall totals, which vary significantly by climate zone. Instead, they relied on empirical cumulative probabilities from local historical distributions. This normalisation converts absolute rainfall into a location-specific measure of relative `extremeness', allowing a consistent hazard assessment in diverse regions. An overview of the variable used in this study is given in table \ref{tab:data_overview}.

Although the dataset goes beyond traditional actuarial information by integrating geocoded environmental and meteorological covariates, it still has important constraints. The proposed framework is intended as a scalable statistical tool for insurers lacking the technical or financial resources to deploy full-scale physical catastrophe models. Consequently, our variables describe local pluvial intensity, rainfall amounts, slope, elevation, and distance metrics, but do not explicitly represent fluvial routing, detailed hydraulic behaviour, fine-scale topographic flow paths, or antecedent soil moisture. Building-level hazard intensity metrics, such as measured inundation depth, are also missing, and preventive actions are only partially reflected in TRI and PPRI indicators. %However, the Bayesian spatial structure offers a flexible baseline that can be incrementally refined as more detailed hydrological or exposure data become accessible.

\begin{table}[!ht]
\centering
\resizebox{\textwidth}{!}{%

\begin{tabular}{llll}
% \hlineB{2}
\cline{2 -4}

 ~~ & \textbf{Variables} & \textbf{Spatial resolution} & \textbf{Comment} \\
% \hlineB{2}
\cline{2 -4}
\textbf{\underline{Insurance}}&  &   &    \\ [0.3cm]

 ~~& claim occurrence, claim amount & building & modelling targets \\ [0.3cm]
 & \makecell[l]{number of rooms, movable assets, \\ precious objects, outbuilding size} & building & underwriting information \\
 
 \textbf{\underline{Climate data}}&  &   &    \\ [0.4cm]

 ~~& \makecell[l]{PPRI, climatic regions,\\ number of CATNAT decrees} & municipality & \makecell[l]{regulatory and historical\\ expert zoning} \\ [0.3cm]
 
 & \makecell[l]{TRI, catchment zones, upwelling areas }& refined polygons & \makecell[l]{regulatory and historical\\ expert zoning} \\
 
 \textbf{\underline{Rainfall data}}&  &   &    \\ [0.4cm]

 ~~& tail weight cluster & municipality &  \makecell[l]{spatial characterisation of\\ rainfall extremes}\\ [0.3cm]
 
 & \makecell[l]{Most Intense Local Rainfall Event\\ (MILRE) and annual MILRE (ann\_MILRE)} & \makecell[l]{ERA5 grid recalculated \\ at the building level} & \makecell[l]{event-based rainfall \\intensity indicators}   \\
 & Water course variables x TRI (WCTRII) & refined polygons & \makecell[l]{interaction between TRI \\and distance/altitude \\to the closest watercourse} \\

 \textbf{\underline{\makecell[l]{Building and \\surrounding data}}}&  &   &    \\ [0.4cm]

 ~~& \makecell[l]{living surface, house value, construction\\ period, number of floors, wall material, \\exterior building surface, total wall \\length, amenity element} & building & \makecell[l]{structural and vulnerability\\ descriptors} \\[0.8cm]
 
 ~~& \makecell[l]{terrain max slope, building density in a \\ 50m buffer, predominant soil type,\\ impervious surface percentage} & \makecell[l]{buffers around \\ the building} & local environmental context \\[0.8cm]

 ~~& \makecell[l]{distance and altitude difference to the \\closest watercourse} & building & \makecell[l]{proximity to hydrographic\\ network}\\

\hline
\end{tabular}}

\caption{Overview of variable used during modelling phase. CATNAT: official natural disaster census \citep{gaspar23}. This table is taken from \citep{Mor25}.}  
\label{tab:data_overview}
    
\end{table}

\section{Models and methods}\label{sec:models}

\subsection{Baseline Generalised Linear Model}

As a baseline specification, we adopt classical generalised linear models (GLMs) that use the complete set of available covariates. Instead of predefined geographic zones, spatial information is explicitly incorporated through detailed building-level, environmental, and hazard-related variables. Since at most one flood claim is observed per policy-year, claim occurrence is modelled with a logistic regression, where the binary outcome $Y_i \in \{0,1\}$ indicates the presence of a claim. Conditional on a claim occurring, the claim size is modelled using a Gamma regression with a log link. Therefore, occurrence and severity are modelled separately \citep{Gol16}. The models are given by: 
\begin{equation}\label{eq:glm}
g(\mathbb{E}(Y_i)) = \beta_0 + \sum_{j=1}^p\beta_j x_{i,j},\end{equation} where $g$ the link function and $\beta$ the coefficients. These baseline models serve as a reference point for assessing spatially enriched specifications. Each policy corresponds to a building geolocated in space and observed over multiple years. In line with common actuarial practice, the temporal dimension is not explicitly modelled. Instead, policy-year observations are treated as independent so that the analysis can focus on the spatial structure of risk \citep{Wah22,Tuf19}. We then incorporate spatial random effects within a Bayesian approach, estimating the resulting models using the \texttt{R-INLA} framework \citep{Rue09,Rue17,Bak18}.

\subsection{Hierarchical Bayesian models for explicit spatial modelling}
In order to explicitly account for spatial dependencies in the modelling framework, we rely on an integrated Bayesian hierarchical model, which for occurrence and severity is of the following form \citep{Rue09}:

\begin{enumerate}
    \item \textbf{Data model (likelihood).}  
    For the occurrence component, because at most one claim is observed per policy-year, the claim count $N_i$ is modelled with a Bernoulli distribution:  
    \begin{equation}\label{eq:BGAM}
    N_i \mid p_i \sim \text{Bernoulli}(p_i), \qquad g(p_i) = \mu + \sum_j f_j(x_{ij}) + \gamma(s_i),
   \end{equation}
    where $g(\cdot)$ denotes the logit link, $f_j$ represent covariate-specific effects, and $\gamma(s_i)$ captures the spatial contribution at location $s_i$.
    
     For the severity component, conditional on a claim occurring ($N_i=1$), the aggregate claim amount $C_i$ is modelled as Gamma-distributed:  
    \[
    C_i \mid \alpha_i, (N_i>0) \sim \text{Gamma}(\alpha_i, \phi), \qquad
    \log(\alpha_i) = \mu + \sum_j f_j(x_{ij}) + \gamma(s_i),
    \]
    \noindent where $\alpha_i = \mathbb{E}(C_i)$ and $\phi=\alpha_i^2/\text{Var}(C_i)$.

    \item \textbf{Latent process.}  
    The covariate effects $f_j(\cdot)$ are specified via Gaussian processes:
    \[
    f_j(x_{ij}) =
    \begin{cases}
    \sum_{v} \beta_{v} \,\mathbf{1}_{\{x_{ij}=v\}}, & \text{if $x_{ij}$ is categorical}, \\
    \sum_{k=1}^L \beta_k \, b_k(x_{ij}), & \text{if $x_{ij}$ is continuous},
    \end{cases}
    \]
    where $b_k(\cdot)$ are the spline basis functions and $\beta_k \sim \mathcal{N}(0, Q^{-1})$.  
    The spatial term $\gamma(s)$ is specified through the discrete or continuous random effect, as detailed in the following.

    \item \textbf{Hyperparameters.} 
    The variance and precision quantities $(\tau, \sigma^2, \kappa, \phi)$ in the latent structure are given prior distributions (hyperpriors). These govern the degree of smoothing, overall scale, and dispersion of the noise.
\end{enumerate}

\noindent This model is called BGAM (Bayesian Generalised Additive Model) in the following. Each $\gamma$ specification corresponds to a different BGAM model that is compared with the baseline GLM. For both claim occurrence and claim severity, the following models are thus defined and evaluated:
\begin{itemize}
    \item \textbf{GLM all}: standard generalised linear model fitted using all available covariates, serving as the baseline model for comparison (Equation \ref{eq:glm}).
    
    \item \textbf{BGAM}: a Bayesian generalised additive model (BGAM) without spatial random effects, used to isolate the contribution of flexible covariate effects (Equation \ref{eq:BGAM} where $\gamma=0$).\vspace{2mm}
    
    \item[] \textbf{Discretely indexed spatial random effects}\vspace{1mm}
    \item \textbf{BGAM + IID}: the BGAM extended with an independent random effect at the municipality level (Equation \ref{eq:BGAM} + Equation \ref{eq:iid}). 
    
    \item \textbf{BGAM + iCAR}: the BGAM that incorporates a structured spatial areal effect specified via an iCAR prior (Equation \ref{eq:BGAM} + Equation \ref{eq:icar}).
    
    \item \textbf{BGAM + BYM}: the BGAM with a BYM spatial component that combines structured and unstructured spatial effects (Equation \ref{eq:BGAM} + Equation \ref{eq:bym}). \vspace{2mm}
    
    \item[] \textbf{Continuously indexed spatial random effects}\vspace{1mm}
    \item \textbf{BGAM + SPDE}: the BGAM includes a continuously indexed spatial random field based on the SPDE representation of a Matérn Gaussian random field (Equation \ref{eq:BGAM} + Equation \ref{eq:spde}).
\end{itemize}

\subsubsection{Discretely indexed spatial random effects}

In this work, spatial aggregation is defined at the municipality level, which is the finest administrative scale commonly used in French insurance practice and in natural disaster reporting. Therefore, each insured is uniquely related to one municipality. Spatial heterogeneity is modelled through municipality-specific latent effects, added to the linear predictors in both the occurrence and severity models. This transforms the baseline model into a mixed-effects model, where spatial variation is represented by random effects indexed by discrete geographic areas. When appropriate, smooth functions of selected covariates are also incorporated. We investigate three standard formulations for discretely indexed spatial random effects, widely employed in actuarial and spatial insurance contexts.

\paragraph*{Independent random effects (IID).}

The most basic specification assumes that the municipality-specific effects are independent and identically distributed:
\begin{equation}\label{eq:iid}
\gamma_j \sim \mathcal{N}(0,\tau^{-1}), \quad j=1,\dots,J,
\end{equation}

where $\tau>0$ is a precision hyperparameter. This setup accounts for unobserved heterogeneity and overdispersion between municipalities, but does not incorporate spatial dependence. Each area is treated independently, with deviations shrunk toward zero. Consequently, the IID model should be viewed as representing purely local random variation rather than genuine spatial correlation.

\paragraph*{Intrinsic conditional autoregressive model (iCAR).}

To explicitly incorporate spatial dependence, we employ the intrinsic conditional autoregressive model (iCAR) \citep{Bes74}, commonly known as the Besag model in \texttt{R-INLA}. Spatial dependence is encoded via an adjacency graph, where two municipalities are defined as neighbours if they share a common border. Let $j \sim i$ indicate that the regions $i$ and $j$ are neighbours, let $\mathbf{v}(i) = \{j : j \sim i\}$ denote the set of neighbours in the region $i$, and let $d_i = |\mathbf{v}(i)|$ be the size of the set. The conditional distribution of the spatial effect in the region $i$ is then given by
\begin{equation}\label{eq:icar}
    \gamma_i \,\mid\, \boldsymbol{\gamma}_{-i}, \tau \;\sim\; 
    \mathcal{N}\!\left( \frac{1}{d_i} \sum_{j \in \mathbf{v}(i)} \gamma_j, \; \frac{1}{d_i \tau} \right),
\end{equation}
where $\tau > 0$ is a precision hyperparameter and $\boldsymbol{\gamma}_{-i}$ denotes the vector of effects for all regions other than $i$.

\noindent In the iCAR specification, the latent effect in each municipality is shifted toward the average of its neighbours, leading to spatial smoothing by borrowing information from adjacent areas. The extent of this smoothing is governed by the single precision parameter $\tau$. This formulation captures broad spatial trends while maintaining local continuity across administrative boundaries.

\paragraph*{Besag-York-Mollié model (BYM).}

Although the iCAR model captures structured spatial variation, it cannot account for unstructured local noise. The Besag-York-Mollié model (BYM) \citep{Bes91} addresses this limitation by decomposing the spatial effect into two terms:
\begin{equation}\label{eq:bym}
\gamma_i = u_i + v_i,
\end{equation}
where $u_i$ has an iCAR prior and $v_i$ is an independent Gaussian random effect. In this way, the BYM formulation enables the model to jointly represent structured spatial variation via $u_i$, encoding correlation between neighbouring areas, and unstructured heterogeneity via $v_i$, capturing location-specific deviations that are not explained by spatial structure.

\medskip

Discretely indexed spatial random effects such as IID, iCAR, and BYM provide an intuitive and interpretable means of introducing spatial structure, which is especially relevant when geographic units have institutional or regulatory importance. In France, the municipal level is particularly meaningful, as natural catastrophe declarations, hazard maps, and pricing schemes frequently use it as the basic spatial unit or are defined at that scale. However, these approaches depend on a predetermined spatial partition and implicitly assume uniform risk within each area. They also treat all adjacency relationships as equally informative, ignoring differences in region size, shape, or internal variability. These shortcomings become particularly problematic for modelling highly local, fine-scale risks. This motivates the adoption of continuously indexed spatial models, discussed in the next section, in which spatial dependence is defined over a continuous domain rather than being tied to administrative boundaries.

\subsubsection{Continuously indexed spatial random effects}\label{sec:spde}
In spatial models defined over a continuous domain, the random effect is specified at each spatial location $s$. Let $\gamma(s)$ denote a spatial random effect for $s \in \mathcal{S} \subset \mathbb{R}^2$. We assume that $\gamma(s)$ is a Gaussian random field (GRF) with a Matérn covariance function,
$$
\mathrm{Cov}\big(\gamma(s), \gamma(s')\big) 
= \sigma^2 \frac{2^{1-\nu}}{\Gamma(\nu)} (\kappa \|s-s'\|)^\nu K_\nu(\kappa \|s-s'\|),
$$
where $\sigma^2$ is the marginal variance, $\kappa$ controls the spatial correlation range, and $\nu$ characterises the smoothness of the field. In applications, $\nu$ is typically fixed because it is only weakly identifiable from insurance data \citep{Lin11}.

Performing direct inference with GRFs is computationally challenging for large datasets, as the associated covariance matrices are dense. The Stochastic Partial Differential Equation (SPDE) method proposed by \citet{Lin11} addresses this by expressing the Matérn GRF as the solution to a linear SPDE. The spatial domain is discretised on a triangulated mesh (commonly built via Delaunay triangulation \citep{Qua94}), and the continuous field is approximated as: 
\begin{equation}\label{eq:spde}
\gamma(s) \approx \sum_{k=1}^M \psi_k(s)\, w_k,
\end{equation}
where $\psi_k(s)$ are piecewise linear basis functions associated with the mesh vertices, and $w_k$ are Gaussian-distributed weights. This yields a Gaussian Markov random field (GMRF) with a sparse precision matrix, facilitating scalable Bayesian inference using INLA. The mesh density governs the balance between the numerical fidelity of the Matérn approximation and computational efficiency.

The INLA-SPDE specification defines spatial dependence on a continuous surface and supports prediction at any location, rather than being limited to a fixed set of areas. This point-level formulation is especially appropriate for flood insurance, where claims are tied to exact building coordinates and risk can differ markedly within the same administrative region. It also lets us explicitly evaluate the added value of latent spatial dependence after incorporating detailed environmental and building-specific covariates, and to contrast continuous spatial models with conventional areal methods.

\subsection{Spatial mesh definition}

France has 34,955 municipalities, of which around 28,000 appear in our portfolio. However, the contracts are distributed very unevenly in space: roughly 85\% of all observations are concentrated in just 6,000 municipalities. This imbalance poses both statistical and computational difficulties for spatial models defined on discrete areal units.

Statistically, specifying a separate random effect for each municipality results in weak parameter identification in areas with little data. In such municipalities, the estimates are mainly governed by prior or spatial smoothing from neighbouring areas, rather than by local exposure, which can lead to excessive shrinkage and unstable inference. On the computational side, areal models require factoring in a $J \times J$ sparse precision matrix, where $J$ denotes the number of spatial units. Although the iCAR and BYM formulations yield sparse structures, the computational burden increases rapidly with $J$ \citep{Bak18}. In the French context, conducting analysis at the municipal level involves managing a much larger number of spatial units than is usual in actuarial studies (for example, compared with the approximately 1,550 spatial tracts examined in the portfolio-level analysis of \citep{Wah22}).
The SPDE approach provides an alternative by modelling space as a continuous domain instead of using predefined administrative regions. The spatial dependence is defined on a triangulated mesh constructed with a constrained refined Delaunay triangulation (CRDT) as implemented in \texttt{R-INLA}. This construction mitigates the occurrence of ill-shaped triangles and enforces boundary constraints such as coastlines. To limit boundary artefacts in the SPDE solution, the mesh is extended beyond the study area with a coarser outer band, which helps stabilise estimation of the range and marginal variance parameters near the domain boundaries.

Mesh resolution is governed by three main hyperparameters: the maximum edge length inside the study region, the maximum edge length in the extended outer region, and the minimum node separation distance (cutoff). These jointly determine the total number of mesh nodes and thus the size and sparsity pattern of the latent precision matrix, with direct implications for both computational expense and spatial resolution.

\begin{figure}[ht]
\centering
\includegraphics[width=0.59\linewidth]{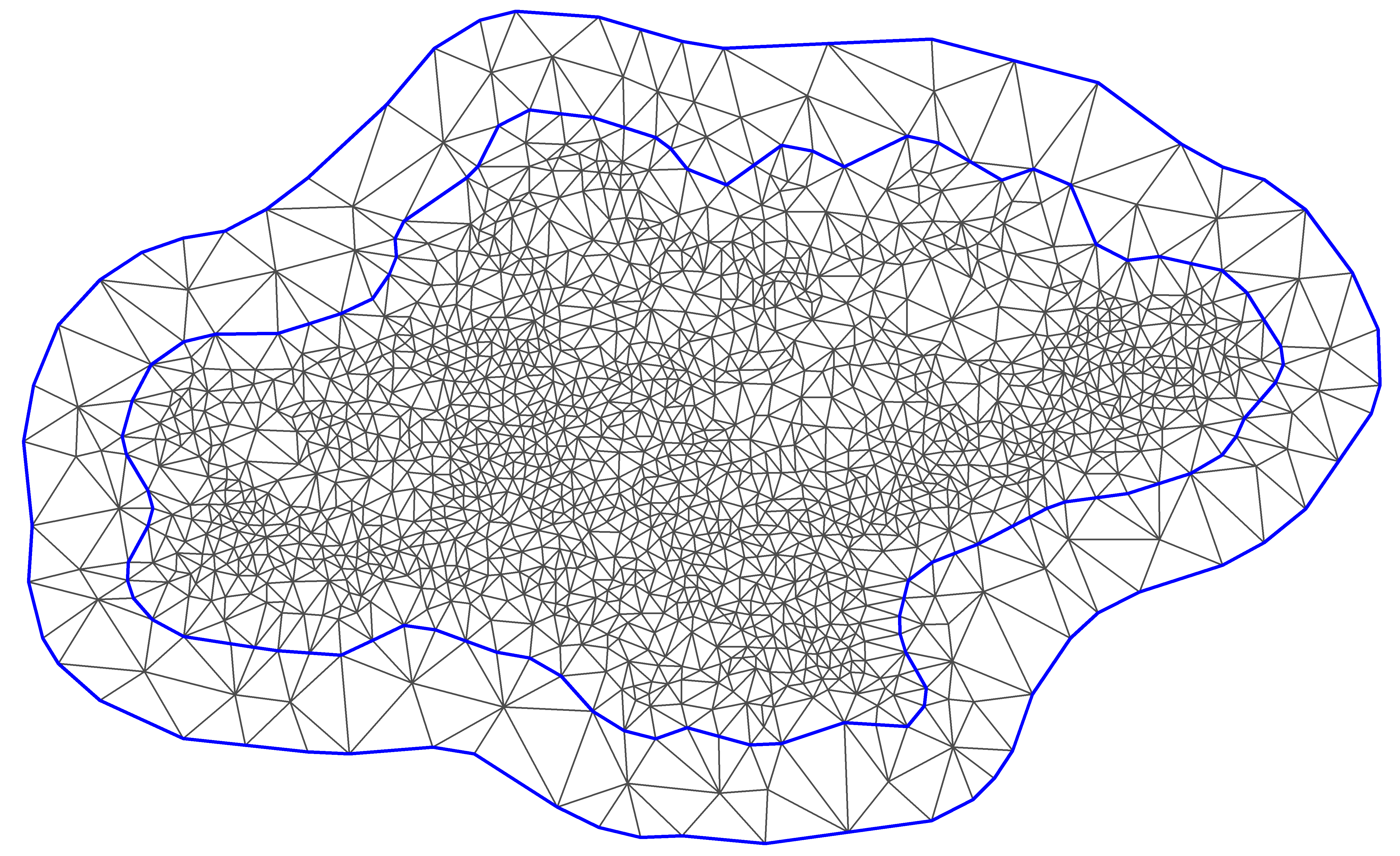}
\caption{SPDE mesh refined in high-density areas represented on the Occitanie region (1{,}512 nodes).}
\label{fig:spde_refined}
\end{figure}

We build a locally refined mesh, characterised by higher resolution in densely populated areas and coarser resolution elsewhere. Hyperparameters are chosen to maximise predictive performance under the constraint that computation time does not exceed that of the discrete areal models. This adaptive strategy permits a more detailed depiction of sub-municipal spatial variability where data are rich, without incurring unnecessary complexity in data-sparse zones. Figure~\ref{fig:spde_refined} shows the mesh constructed over the Occitanie region, comprising 1{,}512 nodes with noticeably dense triangulation in certain zones. Throughout the French territory, the mesh includes 12{,}613 nodes.
The mesh is constructed with a minimum node separation (cutoff) of $0.02^\circ$ (approximately 2~km), a maximum edge length of $1^\circ$ (approximately 110~km) inside the study region, and $1.5^\circ$ in the outer extension. In practice, the inclusion of building locations as mesh vertices produces substantially finer triangulation in data-dense areas, producing the locally refined structure visible in Figure~\ref{fig:spde_refined}.

\subsection{Modelling and evaluation settings}

Bayesian models are fitted using R-INLA. Fixed effects are given zero-mean Gaussian priors with low precision. For IID random effects, we adopt the INLA default priors on the log-precision parameter. Continuous covariates are represented by quadratic B-splines, with the number and placement of knots chosen based on exploratory analysis and expert judgment.

For the SPDE formulation, we employ Penalised Complexity (PC) priors for the Matérn parameters \citep{Fug19}. These PC priors shrink the model toward a baseline with no spatial variation. In practice, we assign exponential priors to the marginal standard deviation $\sigma$ and the spatial range $\rho = \sqrt{8\nu}/\kappa$, and calibrate them so that
$$
\mathbb{P}(\sigma > \sigma_0) = \alpha_1, 
\qquad 
\mathbb{P}(\rho < \rho_0) = \alpha_2,
$$
where the tail probabilities are set to $0.05$ and the thresholds $(\sigma_0,\rho_0)$ are determined using domain knowledge and the spatial scale of the data.
Specifically, we set $P(\rho < 0.02^\circ) = 0.05$ and $P(\sigma > 1) = 0.05$, where distances are expressed in degrees ($0.02^\circ$ corresponds to approximately 2~km at French latitudes). These values encode the prior belief that spatial correlation extends beyond the immediate neighbourhood of each building and that the spatial field remains moderate relative to the covariate contributions. The same prior specification is used for both the occurrence and severity models.

\vspace{0.3cm}
We assess predictive performance on a single validation dataset extracted before any transformations. The validation set consists of policy-years from 2020 to 2022, representing approximately one third of the data, while the training set covers 2014 to 2019. This temporal split ensures chronological separation between training and evaluation, mirroring real-world prediction. The blocked spatio-temporal cross-validation strategy used in \citep{Mor25} was not used here because it is computationally too demanding for INLA-SPDE models, as each fold entails a complete model refitting.
For the severity component, we use the following evaluation metrics: 

\begin{itemize}  
    \item Root Mean Square Error (RMSE) $(\sum_{i=1}^n (\widehat{y}_i - y_i)^2/n)^{1/2}$. This measures the typical size of the prediction errors; smaller RMSE values correspond to a better fit of the model.  \vspace{2mm}

    \item Widely Applicable Information Criterion (WAIC) is a generalisation of the Akaike Information Criterion (AIC) applicable to Bayesian model performance \citep{Wat10}. In INLA it is defined as \citep{Gel14}:
    
     $$\text{WAIC} = -2\underbrace{[\sum_{i=1}^{n} \log\left( \frac{1}{S} \sum_{s=1}^{S} p\left(y_i \mid \theta^{s}\right) \right)}_\text{log pointwise predictive density} - \underbrace{\sum_{i=1}^{n} \mathrm{var}_{\theta}\bigl(\log p(y_i \mid \theta))]}_\text{Effective number of parameter},$$ where $\theta$ represents the posterior distribution of the model and $s$ are samples. Smaller WAIC values signal better performance. The effective number of parameters is used to reflect the complexity of the model. \vspace{2mm}
    
    \item Gini Index assesses the model's ability to rank and differentiate risks, based on the Ordered Lorenz curve of predicted versus realised costs \citep{Den19, Fre14}. We first order policyholders by their predicted risks, such as $\widehat y_{(n)} \geq \dots \geq \widehat y_{(1)}$, then compute cumulative policyholder shares $P_{(i)} = i/n$ and cumulative observed loss shares $R_{(i)} = \frac{\sum_{j=1}^i y_{(j)}}{\sum_{j=1}^n y_j}$.The Gini Index equals twice the area separating the Lorenz curve from the 45-degree equality line.
\end{itemize}  

For classification problems with pronounced class imbalance, accuracy can be misleading \citep{Guo04}. In line with the recommendations of previous work \citep{Jen13, Jap13, Gu09}, instead we employ WAIC for probabilistic calibration, Gini index for ranking quality, and CSI threshold-dependent predictive skill:  

\begin{itemize}  

   \item Critical Success Index (CSI) $\frac{\textit{true positive}}{\textit{true positive}+\textit{false positive}+\textit{false negative}}$ measures how effectively a classifier identifies positive cases in strongly imbalanced datasets \citep{Sch90}. For each model, we select the threshold that maximises CSI in the training set and then report the corresponding CSI in the validation set, along with the associated precision ($\frac{\textit{true positive}}{\textit{true positive}+\textit{false positive}}$) and recall ($\frac{\textit{true positive}}{\textit{true positive}+\textit{false negative}}$) for the minority class. Higher CSI values reflect improved detection of rare events. 
\end{itemize}
\noindent We also report runtimes. All computations were performed on a machine with 88 GB RAM and 16 CPU cores, using 12 cores for parallelisation via the package settings.
All models were fitted using R-INLA version 24.12.11 (R~4.4.2), with the empirical Bayes integration strategy (\texttt{int.strategy = "eb"}), which fixes hyperparameters at their posterior mode rather than integrating over their distribution. This approximation was adopted to maintain computational tractability at the portfolio scale. Exploratory comparisons with fuller integration strategies showed negligible differences in predictive metrics.

\begin{table}[!ht]
\centering
\resizebox{0.95\textwidth}{!}{%
\begin{tabular}{l
                S[table-format=5.0]
                S[table-format=2.1]
                S[table-format=1.1]
                S[table-format=2.2]
                S[table-format=2.2]
                S[table-format=4.0]
                S[table-format=4.0]}
\toprule
Models & {WAIC} & {Gini (\%)} & {CSI (\%)} & {Recall (\%)} & {Precision (\%)} & {Eff. params} & {Runtime (s)} \\
\midrule
\textbf{GLM all}                & 83452 & 55.2 & 3.8 &  7.99 &  6.92 &   \bfseries40 &  \bfseries167 \\
\textbf{BGAM}                    & 83244 & 55.5 & 4.1 &  9.70 &  6.53 &   44 &  421 \\[0.3cm]
\textbf{BGAM + IID}              & 78098 & 62.0 & 6.0 & 10.84 & 11.85 & 3058 & 1350 \\
\textbf{BGAM + iCAR}             & 77556 & 63.6 & 6.8 & \textbf{13.69} & 11.92 & 2469 & 1728 \\
\textbf{BGAM + BYM}              & 77557 & 63.6 & 6.8 & \textbf{13.69} & 11.92 & 2454 & 1851 \\[0.3cm]
% \makecell[l]{\textbf{BGAM + SPDE}\\ \emph{homogeneous mesh}} & 77592 & 62.6 & 6.1 & 11.12 & 13.12 & 1603 &  841 \\
\makecell[l]{\textbf{BGAM + SPDE}\\ \emph{refined mesh}}     & \textbf{63857} & \textbf{66.0} & \textbf{6.9} & 12.30 & \textbf{12.62} & 2477 & 2003 \\
\bottomrule
\end{tabular}}
\caption{Summary of probability models' metrics on the validation set. Recall and precision are for flooded buildings. The effective number of parameters is extracted from the WAIC.}
\label{tab:res_proba_mod}
\end{table}

\section{Results and comparison}\label{sec:results}

\subsection{Flood claim occurrence model}

Table \ref{tab:res_proba_mod} summarises the predictive performance of all occurrence models in the validation set, from which several clear patterns emerge. First, upgrading from standard GLM to BGAM without spatial components yields only modest gains. WAIC and Gini remain almost unchanged, and improvement in CSI is limited. This suggests that once detailed building-level and environmental covariates are taken into account, increasing the flexibility of the functional form alone has little impact on predictive accuracy.

By contrast, the addition of spatial random effects produces a pronounced enhancement across all performance measures and rare-event detection statistics. Even the most basic specification (BGAM + IID) substantially raises the Gini index (from $55.5\%$ to $62.0\%$) and CSI (from $4.07\%$ to $6.00\%$), while also improving both recall and precision for the flooded class. Given the strong class imbalance, this simultaneous increase in recall and precision is especially important: the model correctly flags more flooded buildings while reducing false alarms proportionally. This points to better segmentation in predicted probabilities between flooded and non-flooded buildings.

Within the class of discretely indexed spatial models, performance differences are minor. The iCAR and BYM formulations deliver almost identical predictive metrics, indicating that most spatial dependence is captured through neighbourhood-based smoothing. In light of its lower computational cost and greater conceptual simplicity, the iCAR model is preferable to BYM in this setting. Nevertheless, all discrete spatial specifications entail a large effective number of parameters (on the order of 2,500-3,000), mirroring the high number of municipalities represented in the dataset.

\begin{figure}[ht]
\centering

\begin{subfigure}{0.48\textwidth}
  \centering
  \includegraphics[width=0.95\textwidth]{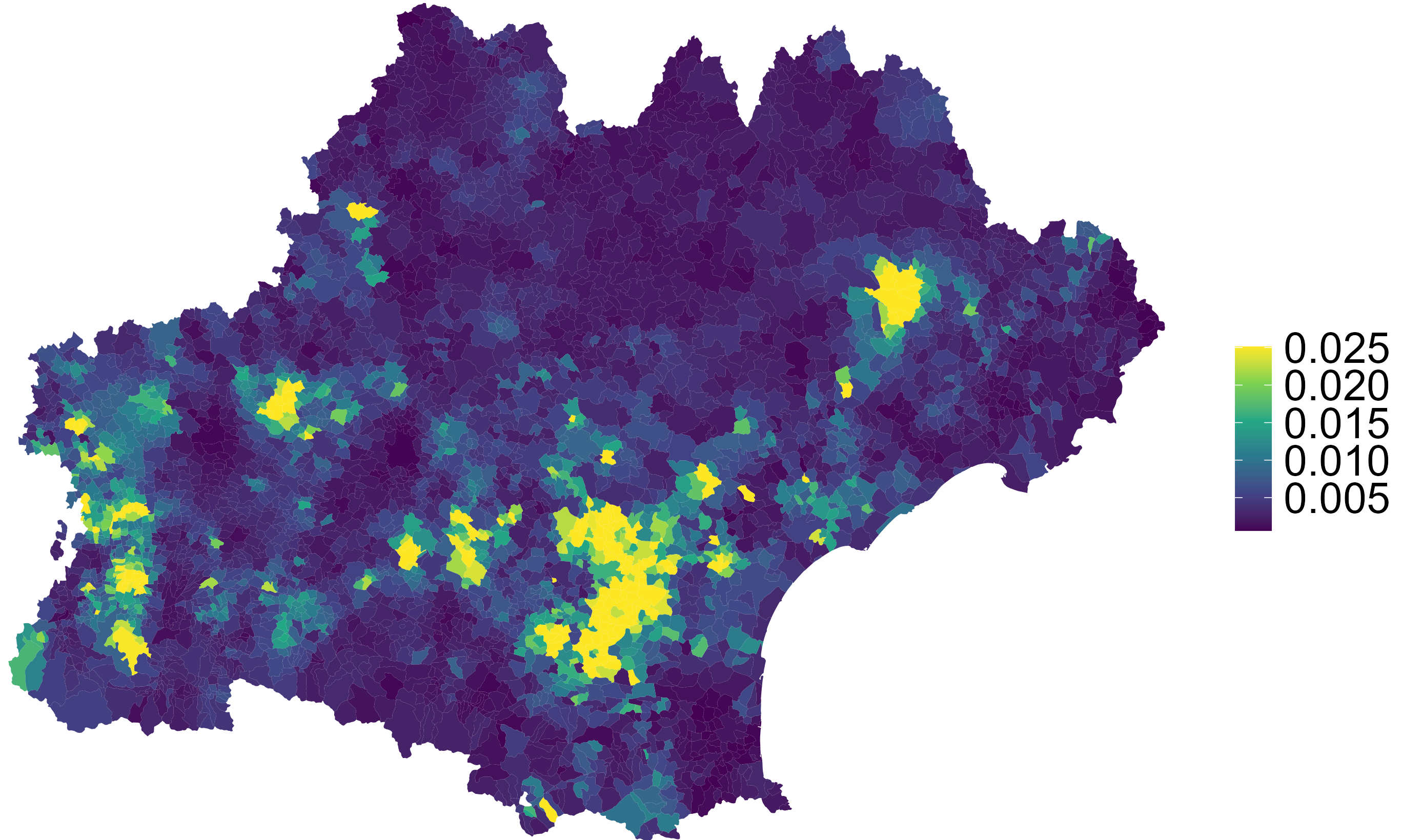}
  \caption{SPDE model.}
  \label{fig:y_pred_spde}
\end{subfigure}
% \hfill
\begin{subfigure}{0.48\textwidth}
  \centering
  \includegraphics[width=0.95\textwidth]{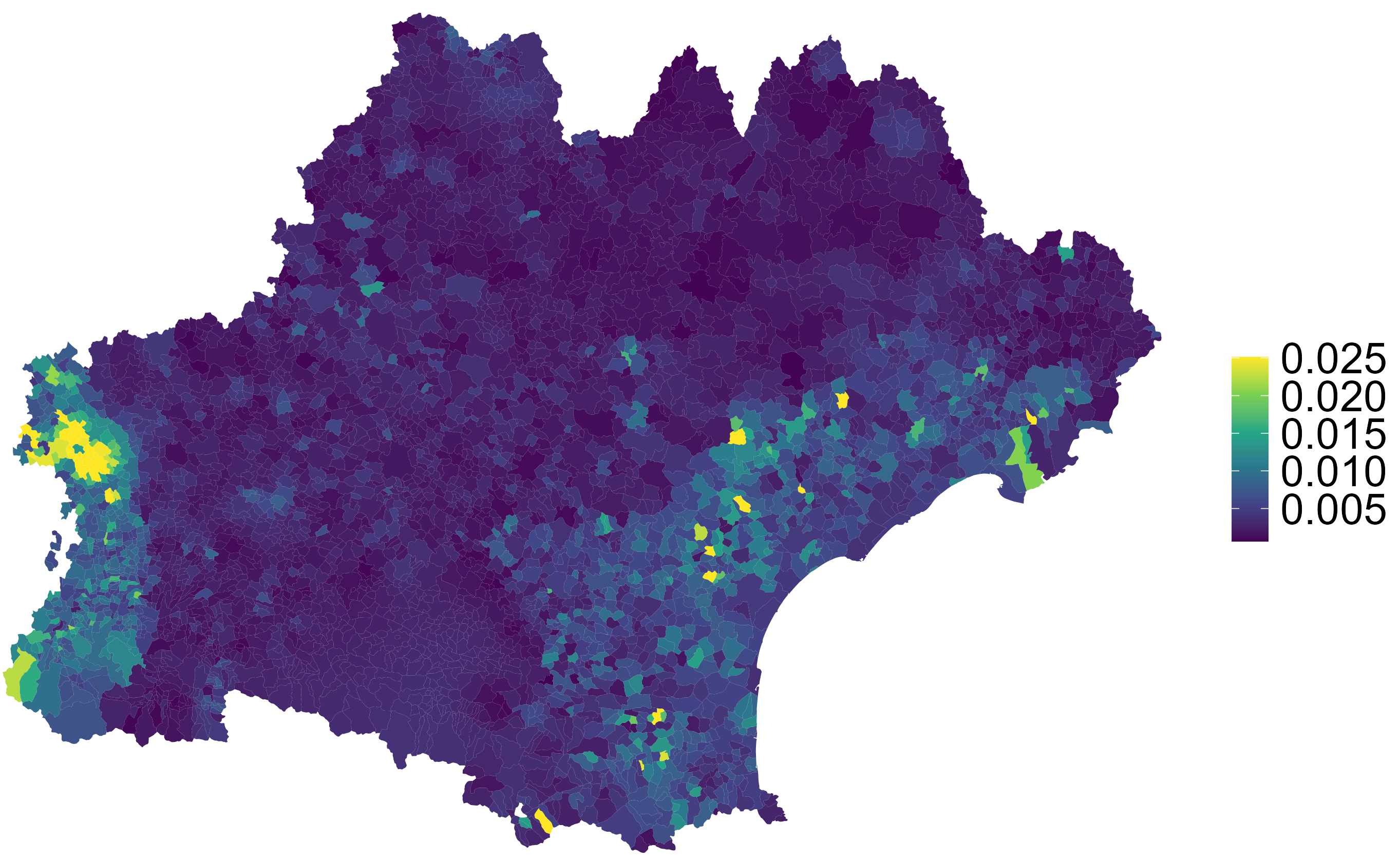}
  \caption{GLM model.}
  \label{fig:y_pred_glm}
\end{subfigure}

\caption{BGAM+SPDE and GLM flood probability of occurrence predictions on the validation set averaged by municipality across policy-years.}
\label{fig:y_pred_occurence}
\end{figure}
\vspace{3mm}

Moving to the continuously indexed model, the SPDE specification outperforms the discrete alternatives in predictive accuracy. The refined mesh specification delivers the strongest overall performance, achieving the lowest WAIC ($63\,857$) and the highest Gini coefficient ($66\%$). The improvement in WAIC is more pronounced than the gain observed with the initial inclusion of spatial effects. Although this specification does not maximise recall, it provides a balanced trade off between recall and precision, with a slightly higher critical success index. Overall, the SPDE framework delivers the strongest predictive performance while allowing greater modelling flexibility through its continuous spatial formulation and adaptable mesh construction.
For comparison, a coarser homogeneous mesh (6,108 nodes) with nearly constant triangle size was also evaluated. Its hyperparameters were chosen so that predictive performance would match that of the best discrete spatial specifications. The resulting model achieved a WAIC of $77\,824$ and a Gini coefficient of $62.6\%$, comparable to the iCAR specification ($77\,556$ and $63.6\%$ respectively), while running in 941 seconds.
\vspace{3mm}

To demonstrate the SPDE specification advantages, we begin by examining aggregated predictions before turning to intra-municipal effects in the dedicated analysis paragraph below. Figure \ref{fig:y_pred_occurence} shows the predicted probabilities of the GLM and the refined BGAM+SPDE model, aggregated at the municipal level for visualisation. Although the predictions are averaged over space, clear structural differences remain visible. At this spatial scale, the outputs of the discrete and continuous models are indistinguishable, so we limit the comparison to the SPDE and the GLM. The differences between the two maps reflect the contribution of the spatial random field, which captures residual spatial structure not explained by the covariates. Regions where the BGAM+SPDE model predicts higher (resp. lower) probabilities than the GLM correspond to positive (resp. negative) spatial effects, highlighting areas where local risk is under or over estimated by the non-spatial specification.

The GLM map highlights broad, large-scale high-risk zones that roughly coincide with areas delineated by institutional hazard maps (e.g., PPRI and TRI), which are among the model covariates. Yet, the GLM does more than reproduce these expert-based classifications. By integrating geolocated building and environmental variables, it generates a risk surface that already shows substantial differentiation within these broader hazard zones. For example, in the southern part of the study area, not all municipalities officially designated as flood-prone exhibit uniformly high predicted risk, underscoring the influence of covariates beyond regulatory maps. The SPDE specification refines this pattern even further. Predicted risks appear less homogeneous within large regions, suggesting that the latent spatial effect is capturing additional local structure not explained by the hazard maps or observed covariates.

Within the SPDE framework, multiple spatially confined areas of elevated risk are revealed, which appear much less distinct or are nearly missing on the GLM map. For example, higher predicted probabilities are found in and around the municipality of Val-d'Aigoual (on the right side of the map, 1,418 inhabitants), an area long known for severe Mediterranean rainfall events and flash floods \citep{PNC20}. Comparable localised patterns arise in parts of the Aude department (toward the center of the map), where small municipalities subject to recurrent river flooding appear more distinctly under the continuous spatial model. Illustrative cases include Saint-Hilaire (713 inhabitants), Molandier (251 inhabitants), and Villegailhenc (1,687 inhabitants). Because these municipalities are small (often with fewer than 2,000 residents), just a few high-risk buildings can result in a high average probability. Moreover, because the economic stakes in these areas are relatively limited, they generally receive less in-depth consideration within comprehensive institutional hazard assessments. Some sites, such as Val-d'Aigoual, have had Flood Prevention Plans mandated since 2002, but they have not been thoroughly analysed or formally validated to date.

\begin{figure}[ht]
\centering
\includegraphics[width=0.99\linewidth]{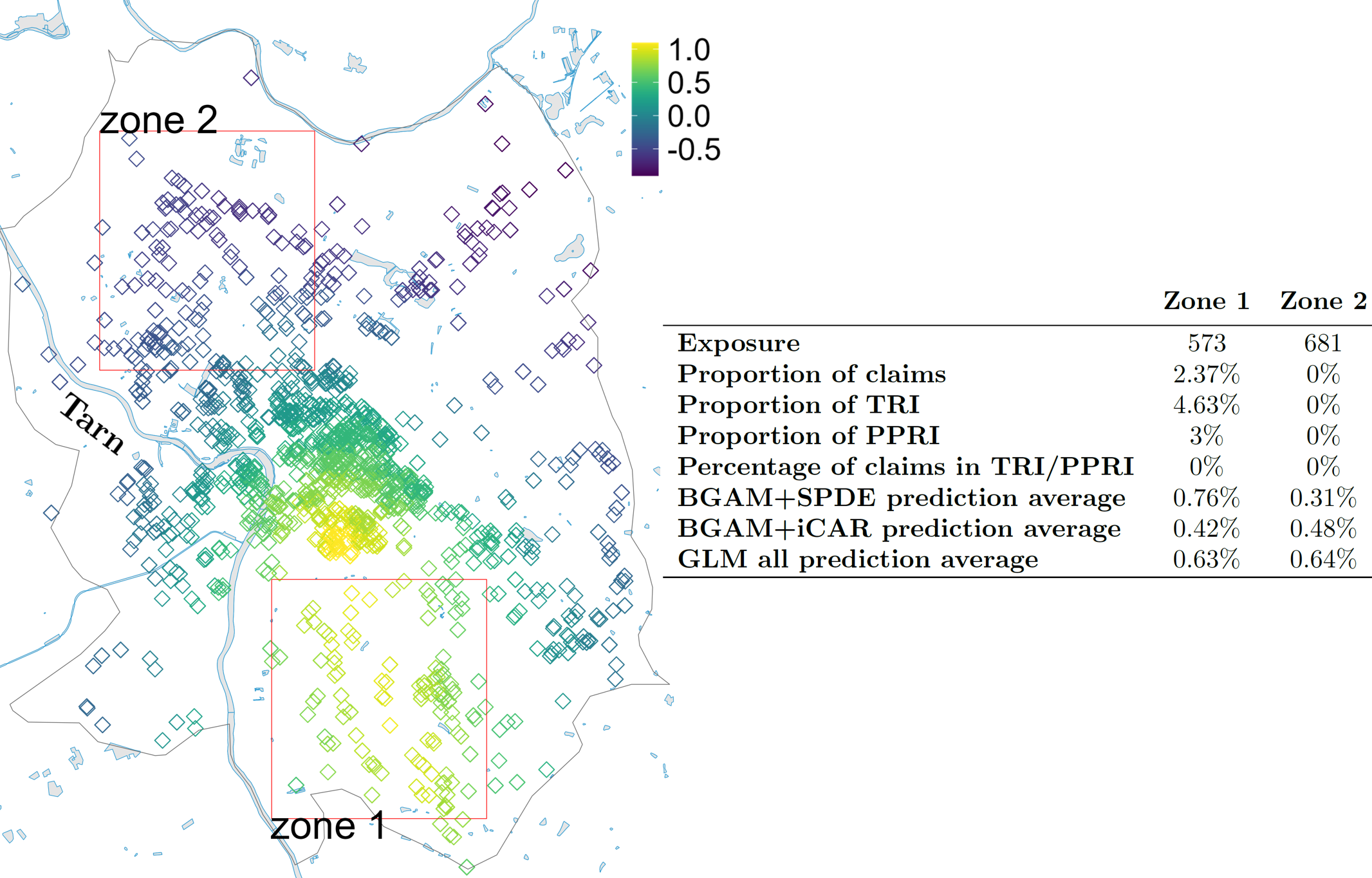}
\caption{Intra-municipal risk heterogeneity in Montauban given by SPDE. Plotted values are BGAM+SPDE building-level estimated spatial effect (Equation \ref{eq:spde} at logit scale), with two subareas (Zone~1 and 2) selected to have similar coverage but contrasting spatial effects. The table reports claims, TRI/PPRI coverage, and mean predicted occurrence probabilities (GLM, BGAM+iCAR, BGAM+SPDE) for each zone.}

\label{fig:example_montauban}
\end{figure}
\paragraph*{Intra-municipality analysis}~~ \\ 

\noindent
Beyond gains in global performance metrics, the continuously indexed SPDE specification provides meaningful intra-municipal risk differentiation. To demonstrate this contribution, we examine Montauban, one of the largest cities in the study area and historically subject to flooding, particularly the Tarn river, which runs through the municipality and has triggered several major overflow events. Figure \ref{fig:example_montauban} shows the estimated spatial effect of SPDE (on logit scale) at the building level, along with two subareas (Zone~1 and Zone~2) chosen to have similar exposure but markedly different spatial effects. This setup highlights the incremental benefit of fine-grained spatial modelling within a single municipality. The associated table summarises the observed results, coverage by institutional flood maps (TRI/PPRI), and average predicted occurrence probabilities for competing models. 

Zone~1, located in the southern part of the city, is covered by official flood-risk instruments: 4.63\% of buildings fall within TRI areas and 2.37\% within PPRI zones. Two flood events were recorded during the study period, yet none of the claims originated from buildings within the TRI/PPRI perimeters. This likely reflects that only the main Tarn watercourse has been considered. It indicates that institutional hazards maps may only partially represent actual building-level flood exposure, either because of their prevention-focused design or due to practical mapping limitations (focus on major assets, available expertise, etc.). 
In comparison, Zone~2 has no official hazard classification and no recorded claims during the study period. At the municipality scale, the GLM and BGAM+iCAR models attribute nearly identical predicted probabilities to the two zones (GLM: $0.63\%$ vs. $0.64\%$; iCAR: $0.42\%$ vs. $0.48\%$). This pattern reflects the role of areal random effects or coarse spatial fields whose signal is essentially constant (or strongly smoothed) within the municipality. For Montauban, the spatial effect of iCAR is estimated at $0.25$ on the logit scale, which changes the overall level of municipal risk, but introduces no variation inside the municipality.

By contrast, the SPDE model generates a marked difference. The mean predicted occurrence probability is 0.76\% in Zone~1 and 0.31\% in Zone~2. For reference, the average flood occurrence in the dataset is 0.22\% at the national level and 0.28\% in Occitanie. Consequently, the SPDE model attributes to Zone~1 a risk more than double the regional mean, while Zone~2 is only modestly higher than that benchmark. This demonstrates how a continuous spatial field can differentiate risk within a single administrative area, rather than uniformly raising or lowering it throughout the municipality.

Taken together, this case illustrates that the continuous SPDE term can uncover sub-municipal residual patterns, highlighting local gradients that are largely smoothed out by zone-based models, while still complementing the geographic information already encoded in external covariates.

\paragraph*{Assessing spatial component influence on other covariates contribution}~~ \\ 

\begin{figure}[ht]
\centering

\begin{subfigure}{0.48\textwidth}
  \centering
  \includegraphics[width=0.95\textwidth]{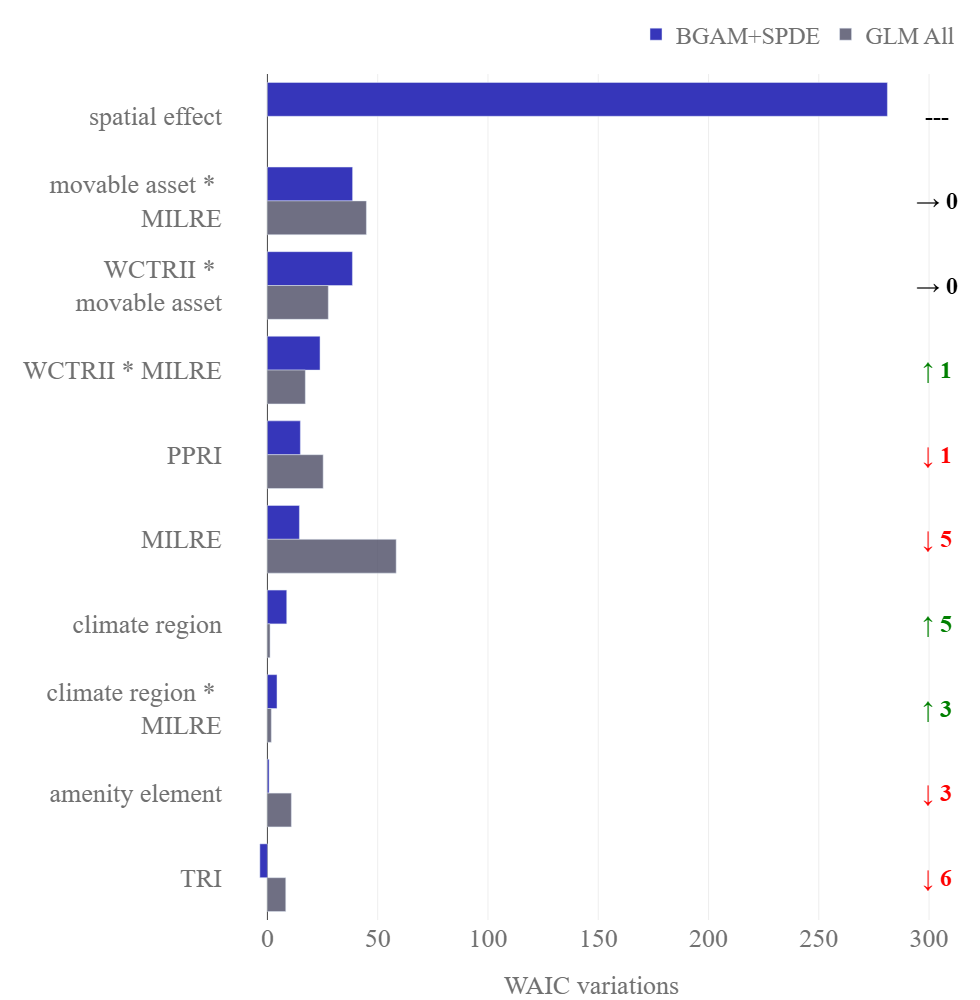}
  \caption{Severity model variable importance variations.}
  \label{fig:imp_cost}
\end{subfigure}
% \hfill
\begin{subfigure}{0.48\textwidth}
  \centering
  \includegraphics[width=0.95\textwidth]{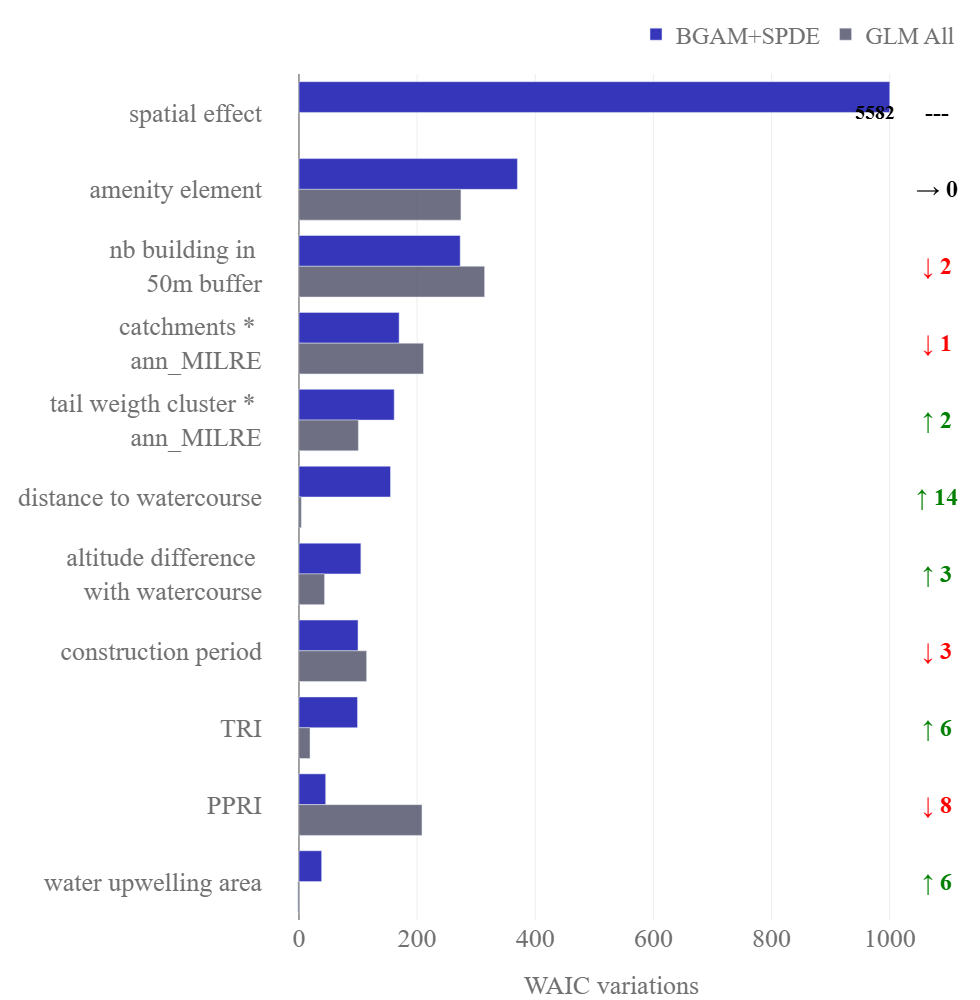}
  \caption{Occurrence model variable importance variations.}
  \label{fig:imp_occ}
\end{subfigure}

\caption{Variable importance analysis for the severity (left) and occurrence (right) models. Importance is measured by the increase in WAIC when a given covariate is removed from the full specification. Larger WAIC differences indicate stronger contribution to model fit. Results are reported for both the GLM All and BGAM+SPDE models. The right-hand side of each panel displays the change in variable ranking between the two specifications, highlighting how the inclusion of an explicit spatial effect modifies the relative importance of covariates.}
\label{fig:two_importance}
\end{figure}

\noindent
To clarify how the explicit spatial term relates to the other predictors, we evaluated the importance of the variable using a WAIC-based strategy. For each covariate, we calculate the change in WAIC between the full model and a reduced model where that covariate is omitted. We implement this procedure for both the GLM and the BGAM+SPDE frameworks. Larger increases in WAIC correspond to a greater contribution to predictive performance. Figure~\ref{fig:two_importance} summarises these importance measures for the occurrence and severity models, ordering the variables by their importance in the BGAM+SPDE specification. In the occurrence model (Figure \ref{fig:imp_occ}), the spatial term is among the most influential factors, underscoring its key role in capturing the spatial structure in the data.

Several large-scale or proxy predictors lose influence once the continuous spatial effect is introduced. In particular, the number of buildings within 50m, a key GLM predictor, becomes less important in the SPDE model. This indicates that the latent spatial field captures the underlying signal of local urban concentration, proximity to water, or construction constraints that building density had been proxying. By directly modelling spatial correlation, the SPDE approach reduces dependence on such proxies and more explicitly represents geographic clustering.

In contrast, variables with a clearer physical interpretation, such as distance to watercourses or detailed hazard indicators (TRI), maintain or increase their relative importance. Institutional zoning indicators like PPRI, typically defined at the municipality level, tend to lose influence in the SPDE setting, indicating that continuous spatial modelling can represent risk patterns more flexibly than coarse administrative categories. These findings show that adding an explicit continuous spatial component not only enhances predictive accuracy but also alters the relative role of the covariates, decreasing dependence on proxy indicators, and emphasising finer-scale environmental variables on flood occurrence. More generally, the latent field is likely capturing physical processes that are not represented in the covariates: local drainage efficiency, smaller watercourses missing from TRI or PPRI, flood protection structures, and particular topographic details. It can therefore be qualitatively associated with concrete physical drivers when combined with local field knowledge.

\subsection{Flood severity model}

\begin{table}[!ht]
\centering
\resizebox{0.85\textwidth}{!}{%
\begin{tabular}{l
                S[table-format=4.0]
                S[table-format=2.1]
                S[table-format=6.0]
                S[table-format=3.0]
                S[table-format=3]}
\toprule
Models & {RMSE} & {Gini (\%)} & {WAIC} & {Eff. params} & {Runtime (s)} \\
\midrule
\textbf{GLM all} & 8835 & 24.7 & 103850 &  \bfseries42 &  \bfseries12 \\
\textbf{BGAM}    & 8745 & 24.8 & 103642 &  37 &  29 \\[0.3cm]
\textbf{BGAM + IID}  & 8694 & 25.0 & 103416 & 776 & 547 \\
\textbf{BGAM + iCAR} & 8669 & 26.8 & 103488 & 627 & 701 \\
\textbf{BGAM + BYM}  & 8663 & 26.8 & 103467 & 622 & 751 \\[0.3cm]
% \makecell[l]{\textbf{BGAM + SPDE}\\ \emph{homogeneous mesh}} & 8674 & 26.7 & 103569 & 440 & 135 \\
\makecell[l]{\textbf{BGAM + SPDE}\\ \emph{refined mesh}}     & \textbf{8656} & \textbf{26.8} & \textbf{103407} & 674 & 358 \\
\bottomrule
\end{tabular}}
\caption{Summary of conditional severity models' metrics calculated on the validation set. The effective number of parameters is extracted from the WAIC.}
\label{tab:res_cost_mod}
\end{table}
Table~\ref{tab:res_cost_mod} summarises the validation results for conditional severity models. Similarly to the occurrence analysis, adding spatial structure consistently enhances predictive performance compared with GLM and non-spatial models. The RMSE drops from 8,835 (GLM) to 8,656 for the refined SPDE specification, while the Gini index increases from 24.7\% to 26.8\%. WAIC values also point to a superior overall fit for spatial models.

However, the magnitude of these gains is smaller than in the occurrence case. Although spatial effects improve performance, the increase in predictive accuracy relative to the added complexity of the model is less substantial. This aligns with previous actuarial studies showing that spatial dependence tends to be more important for claim frequency than for claim severity \citep{Leg24,Tuf19,Wah22}. Severity is generally driven more by building attributes, policy features, and event-specific conditions than by spatial clustering. From a computational standpoint, the differences are also less critical than in the occurrence model because the sample is smaller (only claims are modelled). The runtime increases from 12 seconds for the GLM to 358 seconds for the refined mesh SPDE specification. 

In terms of variable importance (Figure~\ref{fig:imp_cost}), adding the spatial effect leads to only minor changes in ranking relative to the GLM. Unlike in the occurrence model, most covariates preserve similar relative impacts. The main modification concerns rainfall-related predictors, whose importance decreases once the spatial field is included. This indicates that part of the large-scale meteorological signal previously attributed to rainfall covariates is now captured by the spatial random effect. Crucially, the overall contribution of the spatial component to WAIC improvement is much smaller than in the occurrence model. This indicates that residual spatial structure in severity is weaker than in occurrence, and that much of the variation in costs is explained by non-spatial heterogeneity.

\vspace{3mm}
We also investigated whether normalising claim costs by an estimated building value could improve severity modelling, either by modelling a destruction rate or by incorporating the estimate as an offset. However, the building values, imputed from heterogeneous institutional data using a supervised learning model, introduced sufficient measurement error to deteriorate predictive performance under both specifications. Direct cost modelling therefore proved to be the most reliable approach for this portfolio.

\vspace{3mm}

Finally, despite the fact that spatial modelling provides gains for claim cost, its incremental benefit is clearly smaller than for occurrence. The evidence supports the view that spatial dependence is a key determinant of flood occurrence, whereas flood severity is more strongly influenced by individual building attributes and event-specific circumstances.

\subsection{Pure premium analysis}

Although the Montauban case (Figure \ref{fig:example_montauban}) highlights the local benefits of the SPDE specification, pricing accuracy must ultimately be assessed at the portfolio level. We therefore compare the pure premiums generated by the competing occurrence and severity models. For each policy, the pure premium is defined as the product of the predicted probability of occurrence and the predicted conditional loss, following the frequency-severity decomposition \citep{Gol16}.

Rainfall covariates are time-varying and capture the intensity of the event rather than structural exposure. They are used mainly in the literature to explore how portfolio risk responds to different meteorological conditions, as is commonly done in disaster and climate risk modelling \citep{Ass21,Lyu17,Che12}. However, this creates a methodological challenge: rainfall values for future claim events are inherently unobserved at the time of pricing. To obtain an unconditional premium suitable for pricing, we must integrate predictions over the distribution of plausible rainfall scenarios rather than conditioning on any single realisation.

Instead, one could omit the rainfall covariates altogether, resulting in predictions that are, by design, unconditional. However, as demonstrated in \cite{Mor25}, rainfall variables play an important role in improving predictive performance for both occurrence and severity. The Monte Carlo integration described below illustrates how time-varying covariates can be retained for their predictive power while still yielding unconditional premiums that are appropriate for operational pricing.

\vspace{2mm}
\noindent Recall that $\text{ann\_MILRE}$ denotes the maximum annual rainfall (used in the occurrence model) and $\text{MILRE}$ the intensity of the event-specific rainfall (used in the severity model). For a given policy, the pure premium is:
\begin{equation}
\widehat{\pi} = \mathbb{E}_{\text{ann\_MILRE}, \text{MILRE}, \boldsymbol{\theta}}\left[
\mathbb{P}(N>0 \mid \text{ann\_MILRE}, \boldsymbol{\theta}_N) 
\times \mathbb{E}(C \mid N>0, \text{MILRE}, \boldsymbol{\theta}_C)
\right],
\end{equation}
where $\boldsymbol{\theta}_N$ and $\boldsymbol{\theta}_C$ denote the model parameters for occurrence and severity, respectively. This expectation is computed through Monte Carlo integration. For each of the $R$ rainfall scenarios, we sample a complete historical year-event pair $(\text{ann\_MILRE}^{(r)}, \text{MILRE}^{(r)})$ from the observed record, preserving the natural dependence between annual maximum and event-specific rainfall. For each rainfall scenario $r$, we then draw $S$ parameter vectors from the INLA posterior distributions of $\boldsymbol{\theta}_N$ and $\boldsymbol{\theta}_C$, and compute the corresponding probability of occurrence and conditional severities. The pure premium is the average over all $R \times S$ combinations, implementing a discrete approximation of the law of total expectation. We set $R = 1000$ and $S = 100$.

Year-event pairs are resampled with replacement from the nine observed years (2014-2022), preserving the joint dependence between the annual maximum and the event-specific rainfall. For each scenario, posterior parameter vectors are drawn independently for the occurrence and severity models using \texttt{inla.posterior.sample()}, reflecting the separate estimation of the two components. At 100,000 evaluations per policy, the premium estimates are stable; monitoring of running means confirmed convergence well before the full set of draws.
Using 10 cores in parallel, the simulations take about 250 minutes to complete. \vspace{1mm}

This framework jointly accounts for parameter uncertainty and rainfall variability while avoiding any distributional assumptions about future rainfall. Other options include running simulations that rely on climate projections or constructing dependence structures using copulas. We adopt this transparent integration strategy because our primary goal is to compare how different spatial model specifications affect the obtained premium. In Section~\ref{sec:uncertainty}, we additionally use Bayesian predictive distributions to characterise uncertainty and perform simulations. \\ 

To compare the premiums produced by the BGAM+SPDE model with those of the benchmark GLM, we employ ranking-based diagnostics inspired by lift analysis \citep{Gol16,Fre14}. The goal is not only to measure predictive performance but also to examine how each model reallocates risk across the portfolio. Let $\widehat\pi^{\text{SPDE}}_i$ and $\widehat\pi^{\text{GLM}}_i$ denote the predicted pure premiums for policy $i$. We then introduce the pricing relativity index:

% $$
% r_i = \frac{\widehat\pi^{\text{SPDE}}_i}{\widehat\pi^{\text{GLM}}_i}~,
% $$

\begin{equation}
\label{eq:risk_ratio}
\widehat r_i = \frac{\widehat\pi^{\text{GLM}}_i}{\widehat\pi^{\text{SPDE}}_i}~,
\end{equation}

\noindent Values of $r_i < 1$ mean that the SPDE model attributes a higher risk than the GLM, while $r_i > 1$ indicates that the GLM assigns the higher risk. Policies are ordered by $r_i$ and divided into 10 groups with equal total exposure. For each group, we compare the total predicted costs from both models with the total observed losses. Relying on total cost (instead of relative error) ensures that the comparison captures economic relevance at the portfolio level, as various performance measures have already been examined in earlier sections.

\begin{figure}[ht]
  \centering
  \includegraphics[width=0.99\linewidth]{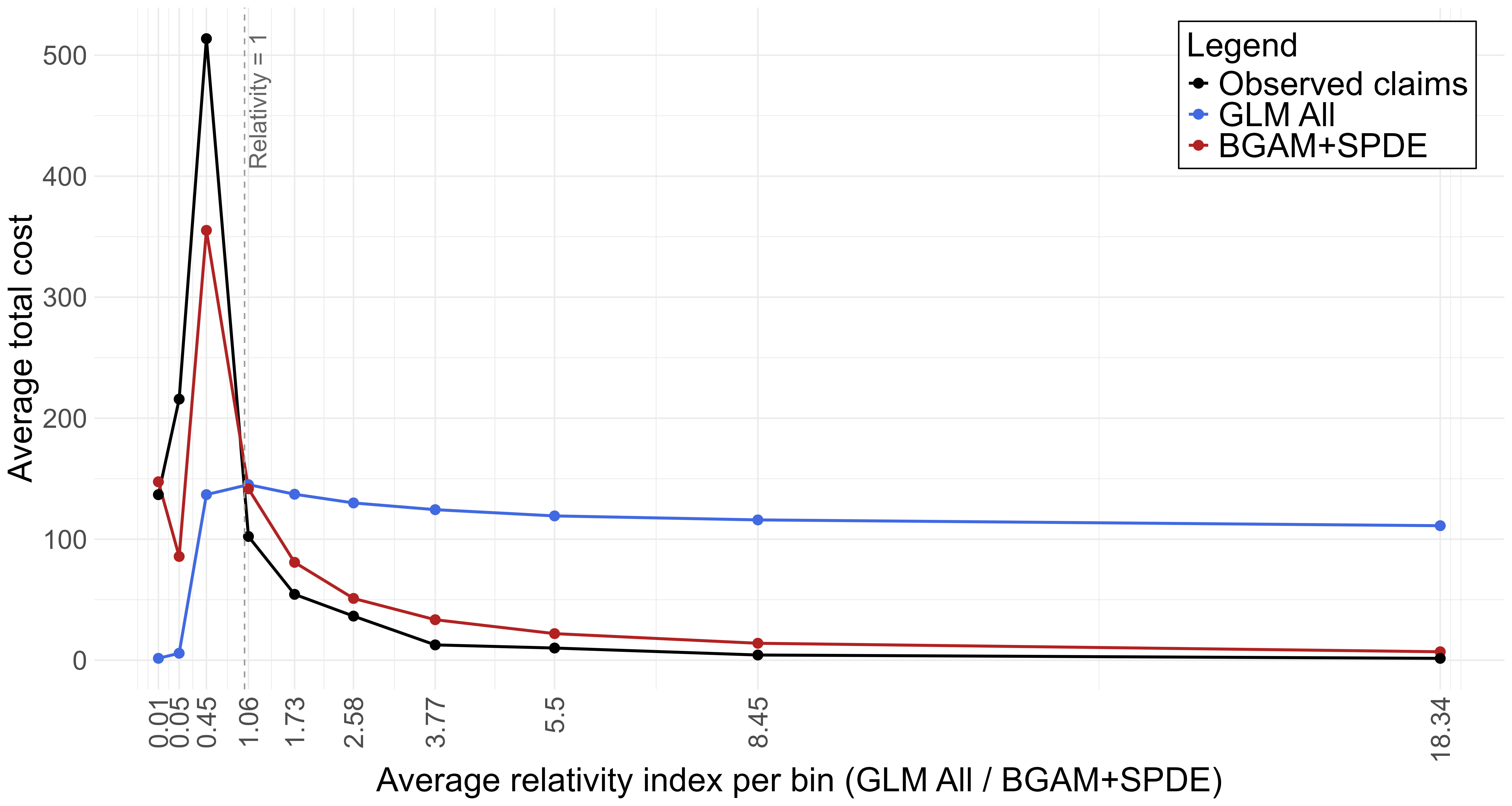}
  \caption{Double lift curve comparing GLM and BGAM+SPDE pure premiums on the validation set. Policies are ordered by the relativity index (Equation \ref{eq:risk_ratio}) and grouped into 10 equal-exposure classes. For each class, average predicted and observed costs are reported. Curves closer to observed losses indicate better risk ranking.}
  \label{fig:double_lift}
\end{figure}

Figure~\ref{fig:double_lift} presents the double lift curves. The GLM systematically overestimates losses in high-relativity segments and underestimates them in low-relativity segments. This pattern is consistent with earlier findings: risks located within broadly defined hazard zones are oversmoothed and consequently overpriced, while localised clusters of elevated risk that are not captured adequately by large-scale covariates tend to be underpriced. In contrast, the BGAM+SPDE curve follows observed losses more closely across all segments, indicating an improved risk ranking and a more accurate allocation of premium volume.

Table~\ref{tab:double_lift_segments} supplements the lift analysis by reporting predicted and realised costs in economically relevant segments. The first panel presents the results by TRI hazard category. Both models yield similar aggregate premiums at the portfolio level, particularly in regions without a TRI designation. However, the differences become more pronounced in the high-risk TRI zones. There, the GLM substantially underestimates the realised costs ($209$K predicted vs $1.4$M observed), while the BGAM+SPDE model sets substantially higher premiums ($707$K). This aligns with the fact that the TRI classification plays a stronger role in the SPDE occurrence model, generating a sharper distinction between TRI levels. This indicates that, even within designated risk zones, the continuous specification helps better explain and capture differences. 

\vspace{3mm}

The second panel considers only municipalities without TRI classification and organises policies according to the relativity index defined in Equation \ref{eq:risk_ratio}.

\noindent This panel enriches the lift curve by providing an absolute value comparison. Recall that low values of $\widehat r_i$ identify policies for which the GLM model assigns substantially lower premiums than the SPDE. In the lowest 10\% segment, the realised losses ($12.3$M) far exceed the predictions of GLM ($1.8$M), while the SPDE model allocates $8.1$M in premium. The concentration is even stronger in the bottom 5\% segment.

At the opposite end of the distribution (where the GLM assigns higher premiums), realised losses remain modest. For example, in the highest 5\% segment, the GLM charges $1.8$M, whereas observed losses are $124$K. This provides quantitative support to the earlier claim that part of the GLM pricing is driven by broad hazard indicators that overgeneralise risk in some areas. This second panel is key because it shows that, although the total predicted premium in non-TRI areas is of comparable magnitude between models, the GLM's premium allocation is markedly less efficient than that of BGAM+SPDE. In summary, the pure premium analysis indicates that our continuous spatial specification enhances portfolio-level risk allocation and further sharpens the identification of localised risk concentrations.

\begin{table}[h]
\centering
\caption{Segment-level comparison of predicted costs and observed outcomes for the BGAM+SPDE and GLM models.}
\label{tab:double_lift_segments}
\resizebox{0.99\textwidth}{!}{%

% \cbstart TABLE 4 UPDATED
\begin{tabular}{lrrrrr}
\toprule
Portfolio segment & $\widehat\pi^{\text{SPDE}}$ & $\widehat\pi^{\text{GLM}}$ & Observed cost & Exposure & Number of claims \\
\midrule
\multicolumn{6}{l}{\textbf{Segmentation by TRI risk level}} \\
\addlinespace
No TRI classification & 17\,546\,772 & 19\,166\,943 & 18\,396\,772 & 1\,312\,865 & 2\,751 \\
TRI - Low risk & 444\,498 & 523\,394 & 317\,951 & 29\,435 & 49 \\
TRI - Medium risk & 1\,037\,065 & 596\,826 & 1\,377\,924 & 40\,309 & 147 \\
TRI - High risk & 706\,845 & 209\,250 & 1\,415\,353 & 11\,791 & 110 \\
\midrule
\multicolumn{6}{l}{\textbf{Segmentation within non-TRI based on $\widehat r = \widehat\pi^{\text{GLM}} / \widehat\pi^{\text{SPDE}}$}} \\
\addlinespace
Lowest 10\% of $\widehat r$ & 8\,058\,583 & 1\,770\,312 & 12\,272\,552 & 130\,958 & 1\,256 \\
Lowest 5\% of $\widehat r$ & 5\,983\,580 & 1\,002\,426 & 10\,243\,673 & 64\,683 & 903 \\
Highest 10\% of $\widehat r$ & 675\,662 & 3\,363\,503 & 269\,705 & 132\,451 & 74 \\
Highest 5\% of $\widehat r$ & 329\,008 & 1\,812\,131 & 123\,801 & 65\,058 & 40 \\
\bottomrule
\end{tabular}}
% \cbend TABLE 4 UPDATED
\end{table}

\subsection{Exploiting Bayesian uncertainty quantification}\label{sec:uncertainty}

\begin{figure}[ht]
  \centering
  \includegraphics[width=0.99\linewidth]{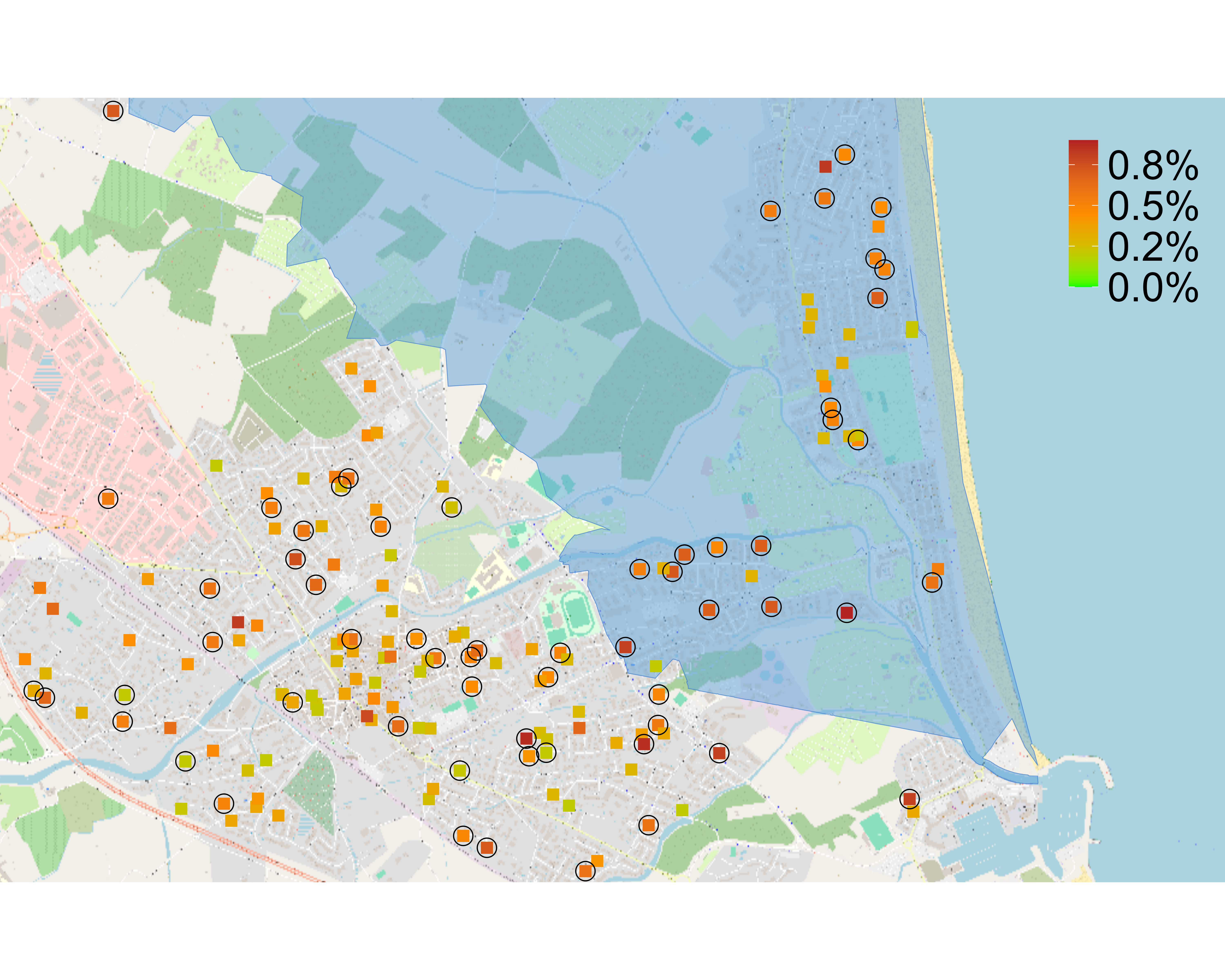}
    \caption{Posterior occurrence probabilities for Argelès-sur-Mer under the BGAM+SPDE model. The map displays predicted mean probabilities, with circled buildings indicating cases where the $97.5\%$ posterior quantile exceeds the municipal average of upper-tail values. The blue overlay denotes the TRI flood risk zone.}

  \label{fig:argelesurmer}
\end{figure}
Beyond improving predictive accuracy, the Bayesian formulation offers a consistent framework to characterise uncertainty. Posterior distributions are obtained for all latent components and predictive quantities, enabling us not only to estimate expected risk levels but also to examine dispersion and upper-tail behaviour. In this subsection, we focus on the $97.5\%$ quantile of the posterior distribution for the probabilities of claim occurrence and demonstrate how this measure can be exploited in an operational setting. For each building in the validation sample, we compute the posterior quantile $97.5\%$ of its predicted claim probability and then identify the municipalities with the highest mean upper-tail values. 

Figure~\ref{fig:argelesurmer} shows the municipality of Argelès-sur-Mer, which displays the highest average posterior $97.5\%$ quantile occurrence probability in the validation portfolio. During the study period, the national disaster database records only a single flood event for this municipality, and no associated claims appear in the training data (the few observed events were assigned to the validation set). Nevertheless, Argelès-sur-Mer is historically recognised as a flood-prone area: it has recorded 12 officially declared natural disasters linked to runoff or river overflow and falls within an approved TRI flood risk zone. Even without any local training claims, the BGAM+SPDE model yields predicted probabilities above the national baseline of roughly $0.22\%$. The risk is not limited to the TRI-designated river overflow area (in blue), but it also affects nearby buildings in a spatially heterogeneous way.

This pattern underlines two central properties of the model. First, predictions are informed by past local claims, but also by building attributes, environmental covariates and rainfall exposure. The spatial random effect further supports extrapolation beyond historically observed events by borrowing information from neighbouring locations. Second, high values of the $97.5\%$ quantile pinpoint the places where extreme outcomes remain statistically credible given the covariate profile and spatial setting. Crucially, the width of the posterior interval (for example, between the $25\%$ and $97.5\%$ quantiles) is not consistently large in the municipality, showing that the highest estimates of the upper-tail arise from structured risk patterns rather than simply from a lack of data. Operationally, municipalities with high predictive values in the upper-tail can justify improved underwriting scrutiny or focused risk assessments, even when recent claims are absent.

\begin{figure*}[t]
  \centering
  \makebox[\textwidth][c]{%
    \includegraphics[width=1.3\textwidth]{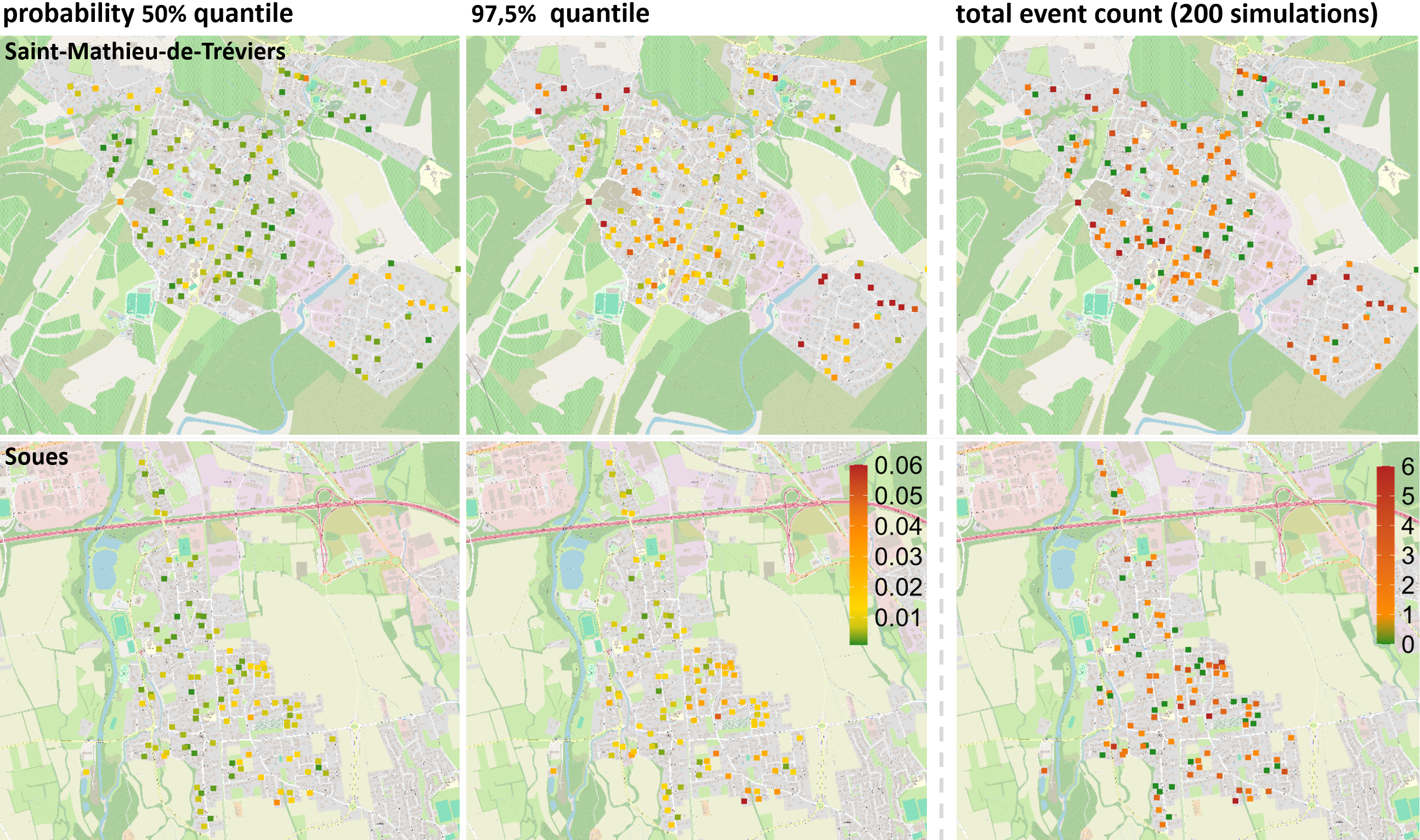}%
  }
  \caption{Posterior predictive distributions for Saint-Mathieu-de-Tréviers and Soues. Maps display the median, $97.5\%$ quantile of simulated occurrence probabilities (200 posterior samples) and total event count per building over the 200 simulated years. Although median risk levels are similar, Saint-Mathieu-de-Tréviers exhibits a substantially heavier upper tail, illustrating spatial heterogeneity in extreme-event uncertainty.}
  \label{fig:simu_soues_saintmat}
\end{figure*}
\vspace{4mm}
The posterior distribution also enables the simulation of claim counts. Using 200 posterior samples of the probability of occurrence, we simulate building-level claim occurrences through Bernoulli trials with these probabilities and then aggregate the resulting claims over the entire portfolio. On the national scale, the simulated median number of claims is $11{,}007$, very close to the total observed of $10{,}800$. For the Occitanie region, the simulated mean is $1{,}633$, again in line with the observed $1{,}426$ claims. These results indicate that the posterior predictive distribution is well calibrated in terms of its central tendency. 

Looking at the upper tail, the $97.5\%$ quantile of simulated national claim counts is $11{,}332$, an $2.34\%$ increase over the simulated median. This moderate spread reflects the assumption of conditionally independent occurrences; within-event spatial correlation, which is not modelled here, would widen the aggregate distribution. In Occitanie, the same quantile is $1{,}728$, or $5.75\%$ above its median. The larger relative spread in Occitanie is consistent with greater exposure to extreme rainfall and a stronger spatial concentration of risk, showing that predictive uncertainty is not uniform across the portfolio. To show how upper-tail uncertainty can distinguish municipalities with comparable average risk, we contrast Saint-Mathieu-de-Tréviers and Soues. These two municipalities are chosen because they have similar median probability of occurrence probabilities but differ in their upper-tail behaviour.

Figure~\ref{fig:simu_soues_saintmat} reports, for each municipality, the median and $97.5\%$ quantile of the distribution of 200 simulated probabilities. The median probabilities are close (around $0.69\%$ for Saint-Mathieu-de-Tréviers and $0.78\%$ for Soues). In contrast, the $97.5\%$ quantile diverges markedly: $2.44\%$ for Saint-Mathieu-de-Tréviers versus $1.49\%$ for Soues. Expressed in terms of simulated annual claim counts under identical exposure, this yields an average of $2.96$ events per year in Saint-Mathieu-de-Tréviers compared to $1.84$ in Soues. Thus, while their average risk levels are comparable, the likelihood of extreme outcomes is substantially higher in Saint-Mathieu-de-Tréviers.

This example highlights a key strength of Bayesian spatial modelling. Regions with similar expected risk can nonetheless exhibit markedly different behaviour in the extremes. From a risk management point of view, municipalities with greater uncertainty in the upper-tail can warrant intensified monitoring, changes in resource allocation, or more conservative underwriting policies. The Bayesian framework with a continuous spatial field offers location-specific posterior uncertainty that reflects local data density and spatial context, capturing heterogeneity in tail behaviour that enriches the distributional output beyond what standard models provide.

\section{Conclusion}

This paper demonstrates the benefits of explicitly accounting for spatial dependence when pricing flood risk in homeowners insurance. Using a high-resolution portfolio enriched with building-level and meteorological data, we benchmarked classical GLMs with a hierarchy of Bayesian spatial models. We find that incorporating spatial structure significantly enhances occurrence prediction and portfolio risk ranking, with the SPDE formulation offering the best balance between predictive performance, computational cost, and interpretability at sub-municipal scales. This point-level specification captures local risk gradients that remain hidden in areal models.

The Bayesian approach not only refines point estimates but also provides informative measures of uncertainty. Analysis of the upper tail uncovers variation in extreme-event risk that mean-based predictions fail to detect, flagging municipalities where enhanced underwriting scrutiny or targeted capital deployment may be appropriate. Relative to conventional zoning approaches or residual-based adjustments, a continuously indexed spatial model limits oversmoothing across major hazard regions while uncovering small-scale clusters of elevated risk that are crucial for climate-exposed insurance portfolios.

Although INLA provides a flexible alternative to full Markov Chain Monte Carlo, its computational burden is still substantial. Even with high‑performance hardware and parallelisation, fitting the most complex models was time‑consuming, limiting iterative research and operational use. To preserve tractability, we constrained model complexity by restricting extra spline components and random effects and by adopting simplified integration schemes in INLA. Runtime seemed driven mainly by model architecture, especially the latent field dimension and number of hyperparameters, rather than by sample size. Future research could aim to preserve the advantages of continuous spatial modelling while reducing computational overhead, for instance through more scalable approximations or hybrid approaches combining spatial statistics and machine learning.

From an actuarial standpoint, introducing continuous spatial effects has significant consequences for both pricing and risk pooling. Compared to the GLM framework, the SPDE specification reduces broad mutualisation and yields more localised risk differentiation. As a result, premiums are more closely aligned with actual loss experience and risk ranking improves, especially in segments where the benchmark model heavily understates losses. For example, in the lowest $10\%$ of the relativity index, the realised losses total $12.3$M, while the GLM assigns only $1.8$M in premiums; the SPDE model reallocates $8.1$M, moving much closer to the observed losses while still preserving some pooling.

This highlights both the benefits and the limits of spatial refinement. Finer-grained modelling enables more accurate and actuarially fair pricing, more efficient capital allocation, and potentially better targeting of prevention efforts. At the same time, reduced mutualisation can widen premium differences among nearby policyholders and trigger affordability or regulatory concerns in highly exposed regions, thus adversely affecting a notion of societal fairness. As a result, although spatial models are a crucial instrument for technical risk management, their deployment must be carefully calibrated to broader considerations of market competition, social acceptability, and the changing context of climate risk mitigation.

\section*{Acknowledgement(s)}

We thank ADDACTIS France for granting access to the insurance data used in this study, and Franck Baton (climate scientist) for his valuable discussions on flood modelling.

% \section*{Disclosure statement}

% An unnumbered section, e.g.\ \verb"\section*{Disclosure statement}", may be used to declare any potential conflict of interest and included \emph{in the non-anonymous version} before any Notes or References, after any Acknowledgements and before any Funding information.

\section*{Funding}

This study was not supported by any particular grant from funding bodies in the commercial, governmental, or non-profit sectors.

% \section*{Notes on contributor(s)}

% An unnumbered section, e.g.\ \verb"\section*{Notes on contributors}", may be included \emph{in the non-anonymous version} if required. A photograph may be added if requested.

% \section*{Nomenclature/Notation}

% An unnumbered section, e.g.\ \verb"\section*{Nomenclature}" (or \verb"\section*{Notation}"), may be included if required, before any Notes or References.

% \section*{Notes}

% An unnumbered `Notes' section may be included before the References (if using the \verb"endnotes" package, use the command \verb"\theendnotes" where the notes are to appear, instead of creating a \verb"\section*").

\section*{Data Availability Statement}
The raw data employed to construct the variables, as well as the French disaster database (GASPAR), are predominantly open-source datasets produced by French public institutions and can be accessed at \url{https://www.data.gouv.fr/}. For reasons of confidentiality, the insurance portfolio used in this analysis cannot be made publicly available.
% \section{References}

\section{Appendices}
\appendix

\section{Rainfall Related Variable Definitions}\label{app:rainfall}

This appendix outlines the definitions of the three primary meteorological variables used in our modelling framework, following the methodology established in \cite{Mor25}: \texttt{tail\_weight\_cluster}, \texttt{MILRE}, and \texttt{ann\_MILRE}. These indicators were specifically developed to address the spatial non-stationarity of rainfall extremes and the multi-scale nature of flood-triggering events.

\subsection*{Tail Weight Cluster}
To represent the wide range of climatic conditions across France, the asymptotic properties of daily rainfall are described using Extreme Value Theory. The approach involves fitting a Generalised Pareto Distribution (GPD) to exceedances over high, location-specific thresholds at each grid cell. The estimated shape parameter serves as a metric for tail heaviness, capturing the local susceptibility to extreme precipitation. To provide a compact spatial representation for predictive modelling, grid cells with similar tail characteristics are grouped through clustering. The resulting categorical feature, \texttt{tail\_weight\_cluster}, allows the model to distinguish between regions characterised by heavy-tailed extremes, such as the Mediterranean south, and those with milder northern climates.

\subsection*{The Most Intense Local Relative Event (MILRE)}
The \texttt{MILRE} indicator was designed to reconcile spatial heterogeneity with the varying temporal scales associated with flood-generating rainfall. For a given event, the cumulative rainfall $i_{nd}^*$ is calculated over multiple durations $nd \in \{1, 3, 5, 7, 10, 30\}$ days. This set of durations is chosen to encompass a broad spectrum of hydrological triggers, ranging from short-lived flash-flood conditions (1-3 days) to longer-term accumulation and soil saturation (10-30 days). To reflect local exposure, each insured building is assigned the maximum value among its four nearest ERA5-Land grid cells.

A fundamental challenge in regional flood modelling is the lack of comparability between absolute rainfall totals across different climates; for example, 50 mm of precipitation may be unremarkable in one region while representing a catastrophic extreme in another. To standardise these values across locations, the absolute depths are converted into empirical cumulative probabilities $\widehat{F}_{nd}$ derived from the local historical distribution.
Empirical CDFs are computed over the period 1950 to 2013, using only data prior to the insurance observation window (2014-2022) to avoid look-ahead bias.
The normalisation is computed as:
$$
    \widehat F_{nd}(i_{nd}^*) = \frac{1}{M_{nd}}\sum_{k=1}^{M_{nd}} \mathbbm{1}_{i_{nd}^k \leqslant i_{nd}^*},
$$
where $M_{nd}$ is the length of the historical series in days. This normalisation yields a relative measure of extremity that is directly comparable across geographically diverse regions. The \texttt{MILRE} is defined as the maximum of these empirical probabilities across all temporal windows:
$$
    MILRE = \max_{nd \in \{1, 3, 5, 7, 10, 30\}} \widehat F_{nd}(i_{nd}^*).
$$
By applying the max-operator over these windows, the indicator remains independent of a specific flood typology, automatically highlighting the most significant meteorological stress relative to the local climatology.

\subsection*{Annual Aggregate for Occurrence Modelling (\texttt{ann\_MILRE})}
For modelling of claim frequency and occurrence, a single event date is often insufficient because insurance policies cover durations (e.g., one year) during which significant exposure may occur without resulting in a claim. Consequently, an annual version, \texttt{ann\_MILRE}, is used. This variable is constructed using the same principles as the \texttt{MILRE}, with the distinction that it identifies the maximum accumulated value for each time window over an entire year rather than at a discrete event date. This captures the most extreme relative event experienced by a property during the policy year, ensuring that the occurrence models incorporate a robust predictor of meteorological pressure even in the absence of observed claims.


\begin{thebibliography}{}




% \bibitem[American Academy of Actuaries(2018)]{AAA18}  % uncited – commented out
% American Academy of Actuaries, Extreme Events and Property Lines Committee. Uses of catastrophe model output (2018). Actuary{.}org.

\bibitem[André et~al.(2013)André, Monfort, Bouzit \& Vinchon]{And13}
André, C., Monfort, D., Bouzit, M., \& Vinchon, C. (2013). Contribution of insurance data to cost assessment of coastal flood damage to residential buildings: insights gained from Johanna (2008) and Xynthia (2010) storm events. Natural Hazards and Earth System Sciences, 13(8), 2003-2012.

\bibitem[Assunção et~al.(2014)Assunção, Costa, Prates \& Silva e Silva]{Ass14}
Assunção, R., Costa, M. A., Prates, M. O., \& Silva e Silva, L. S. G. (2014). Spatial analysis. In A. Charpentier (Ed.), Computational actuarial science with R (pp. 207-256). Chapman \& Hall.

\bibitem[France Assureurs(2021)]{Ass21}
France Assureurs. (2021). Impact du changement climatique sur l'assurance à l'horizon 2050. Étude FA.

\bibitem[Bakka et~al.(2018)]{Bak18}
Bakka, H., Rue, H., Fuglstad, G. A., Riebler, A., Bolin, D., Illian, J., ... \& Lindgren, F. (2018). Spatial modelling with R‐INLA: A review. Wiley Interdisciplinary Reviews: Computational Statistics, 10(6), e1443.

\bibitem[Bernet et~al.(2019)]{Ber19}
Bernet, D. B., Trefalt, S., Martius, O., Weingartner, R., Mosimann, M., Röthlisberger, V., \& Zischg, A. P. (2019). Characterizing precipitation events leading to surface water flood damage over large regions of complex terrain. Environmental Research Letters, 14(6), 064010.

\bibitem[Besag et~al.(1991)Besag, York, \& Mollié]{Bes91}
Besag, J., York, J., \& Mollié, A. (1991). Bayesian image restoration, with two applications in spatial statistics. Annals of the institute of statistical mathematics, 43(1), 1-20.

\bibitem[Besag(1974)]{Bes74}
Besag, J. (1974). Spatial interaction and the statistical analysis of lattice systems. Journal of the Royal Statistical Society: Series B (Methodological), 36(2), 192-225.

\bibitem[Boa, Underwood, \& Wilkins(2006)]{Boa06}
Boa, J.M., Underwood, A.M. and Wilkins, W.R. (2006). Casualty Actuarial Society. In Encyclopedia of Actuarial Science (eds J.L. Teugels, B. Sundt and J. Lemaire).

\bibitem[Boskov \& Verrall(1994)]{Bos94}
Boskov, M., \& Verrall, R. J. (1994). Premium rating by geographic area using spatial models. ASTIN Bulletin: The Journal of the IAA, 24(1), 131-143.


\bibitem[Boudreault \& Ojeda(2022)]{Bou22}
Boudreault, M., \& Ojeda, A. (2022). Ratemaking territories and adverse selection for flood insurance. Insurance: Mathematics and Economics, 107, 349-360.


\bibitem[CEPRI(2022)]{Cep22}
Centre Européen de Prévention du Risque Inondation. (2022). French implementation of the European Flood Directive. CEPRI. \url{https://cepri.net/les-outils-a-votre-disposition/les-lois-et-le-risque-inondation/loi-grenelle-2/}

\bibitem[Chatelain \& Loisel(2021)]{Cha21}
Chatelain, P., \& Loisel, S. (2021). Subsidence and household insurances in France: geolocated data and insurability.


\bibitem[Cheng et~al.(2012)Cheng, Li, Li, \& Auld]{Che12}
Cheng, C. S., Li, Q., Li, G., \& Auld, H. (2012). Climate change and heavy rainfall-related water damage insurance claims and losses in Ontario, Canada. Journal of Water Resource and Protection, 4(2), 49-62.

\bibitem[Denuit et~al.(2019)Denuit, Sznajder, \& Trufin]{Den19}
Denuit, M., Sznajder, D., \& Trufin, J. (2019). Model selection based on Lorenz and concentration curves, Gini indices and convex order. Insurance: Mathematics and Economics, 89, 128-139.

\bibitem[Denuit(2004)]{Den04}
Denuit, M., \& Lang, S. (2004). Non-life rate-making with Bayesian GAMs. Insurance: Mathematics and Economics, 35(3), 627-647.

% \bibitem[Denuit et~al.(2004b)Denuit, Charpentier, \& Bébéar]{Den04b}  % uncited – commented out
% Denuit, M., Charpentier, A., \& Bébéar, C. (2004). Mathématiques de l'Assurance Non-Vie. Tome I: Principes Fondamentaux de Théorie du Risque.

\bibitem[Dimakos \& Di Rattalma(2002)]{Dim02}
Dimakos, X. K., \& Di Rattalma, A. F. (2002). Bayesian premium rating with latent structure. Scandinavian Actuarial Journal, 2002(3), 162-184.


% \bibitem[Dorfman(1979)]{Dor79}  % uncited – commented out
% Dorfman, R. (1979). A formula for the Gini coefficient. The review of economics and statistics, 146-149.

\bibitem[DRIEAT(2023)]{Dri23}
DRIEAT (Direction régionale et interdépartementale de l'environnement, de l'aménagement et des transports). (2023). Plan de prévention des risques inondation. Ministère de la Transition écologique. \url{https://www.ecologie.gouv.fr/politiques-publiques/prevention-inondations}


% \bibitem[Dutta(2025)]{Dut25}  % uncited – commented out
% Dutta, S., van Niekerk, J., \& Rue, H. (2025). Scalable skewed Bayesian inference for latent Gaussian models. arXiv preprint arXiv:2502.19083.


\bibitem[ECMWF(2019)]{ECMWF19}
Muñoz Sabater, J. (2011). ERA5-Land hourly data from 1950 to present. Copernicus Climate Change Service (C3S) Climate Data Store (CDS). %10.24381/cds.e2161bac

\bibitem[Emanuelsson(2011)]{Ema11}
Emanuelsson, P. (2011). Construction of rating territories for water-damage claims (Doctoral dissertation, Stockholm University).


% \bibitem[Ferkingstad(2015)]{Fer15}  % uncited – commented out
% Ferkingstad, E., \& Rue, H. (2015). Improving the INLA approach for approximate Bayesian inference for latent Gaussian models.

% \bibitem[France Assureurs(2021)]{Franceass21}  % duplicate of Ass21 – commented out
% {France Assureurs} (2021). Impact du changement climatique sur l'assurance \`a l'horizon 2050. \'Etude FA.


\bibitem[Frees et~al.(2014)Frees, Meyers, \& Cummings]{Fre14}
Frees, E. W., Meyers, G., \& Cummings, A. D. (2014). Insurance ratemaking and a Gini index. Journal of Risk and Insurance, 81(2), 335-366.


\bibitem[Fuglstad et~al.(2019)Fuglstad, Simpson, Lindgren, \& Rue]{Fug19}
Fuglstad, G. A., Simpson, D., Lindgren, F., \& Rue, H. (2019). Constructing priors that penalize the complexity of Gaussian random fields. Journal of the American Statistical Association, 114(525), 445-452.

% \bibitem[Gaedke-Merzhäuser et~al.(2023)Gaedke-Merzhäuser, Krainski, Janalik, Rue, \& Schenk]{Gae23}  % uncited – commented out
% Gaedke-Merzhäuser, L., Krainski, E., Janalik, R., Rue, H., \& Schenk, O. (2023). Integrated nested laplace approximations for large-scale spatial-temporal bayesian modelling. arXiv preprint arXiv:2303.15254.

\bibitem[GASPAR(2023)]{gaspar23}
GASPAR, Ministère de la Transition écologique. Base nationale de Gestion ASsistée des Procédures Administratives relatives aux Risques (GASPAR). Portail data.gouv. {\sloppy\url{https://www.data.gouv.fr/fr/datasets/base-nationale-de-gestion-assistee-des-procedures-administratives-relatives-aux-risques-gaspar}\par}

\bibitem[Gelman et~al.(2014)Gelman, Hwang, \& Vehtari]{Gel14}
Gelman, A., Hwang, J., \& Vehtari, A. (2014). Understanding predictive information criteria for Bayesian models. Statistics and computing, 24(6), 997-1016.

\bibitem[Goldburd et~al.(2016)Goldburd, Khare, Tevet, \& Guller]{Gol16}
Goldburd, M., Khare, A., Tevet, D., \& Guller, D. (2016). Generalized linear models for insurance rating. Casualty Actuarial Society, CAS Monographs Series, 5, 77.

% \bibitem[Gradeci et~al.(2019)Gradeci, Labonnote, Sivertsen, \& Time]{Gra19}  % uncited – commented out
% Gradeci, K., Labonnote, N., Sivertsen, E., \& Time, B. (2019). The use of insurance data in the analysis of Surface Water Flood events-A systematic review. Journal of Hydrology, 568, 194-206.


\bibitem[Grahn \& Nyberg(2014)]{Gra14}
Grahn, T., \& Nyberg, R. (2014). Damage assessment of lake floods: Insured damage to private property during two lake floods in Sweden 2000/2001. International Journal of Disaster Risk Reduction, 10, 305-314.

\bibitem[Gschlößl \& Czado(2007)]{Gsc07}
Gschlößl, S., \& Czado, C. (2007). Spatial modelling of claim frequency and claim size in non-life insurance. Scandinavian Actuarial Journal, 2007(3), 202-225.

\bibitem[Gu et~al.(2009)Gu, Zhu, \& Cai]{Gu09}
Gu, Q., Zhu, L., \& Cai, Z. (2009, October). Evaluation measures of the classification performance of imbalanced data sets. In International symposium on intelligence computation and applications (pp. 461-471). Berlin, Heidelberg: Springer Berlin Heidelberg.

\bibitem[Guo(2004)]{Guo04}
Guo, H., \& Viktor, H. L. (2004). Learning from imbalanced data sets with boosting and data generation: the databoost-im approach. ACM Sigkdd Explorations Newsletter, 6(1), 30-39.

% \bibitem[He(2009)]{He09}  % uncited – commented out
% He, H., \& Garcia, E. A. (2009). Learning from imbalanced data. IEEE Transactions on knowledge and data engineering, 21(9), 1263-1284.

\bibitem[Illian et~al.(2012)Illian, Sørbye, \& Rue]{Ill12}
Illian, J. B., Sørbye, S. H., \& Rue, H. (2012). A toolbox for fitting complex spatial point process models using integrated nested Laplace approximation (INLA).

\bibitem[INSEE(2024)]{Insee24}
Institut National de Statistiques et des Etudes Economiques (2024). Insee analyses Occitanie. Self-Published Report. Retrieved from \url{https://www.insee.fr/fr/statistiques/8264502}

\bibitem[Japkowicz(2013)]{Jap13}
Japkowicz, N. (2013). Assessment metrics for imbalanced learning. Imbalanced learning: Foundations, algorithms, and applications, 187-206.

\bibitem[Jeni et~al.(2013)Jeni, Cohn, \& De La Torre]{Jen13}
Jeni, L. A., Cohn, J. F., \& De La Torre, F. (2013, September). Facing imbalanced data--recommendations for the use of performance metrics. In 2013 Humaine association conference on affective computing and intelligent interaction (pp. 245-251). IEEE.

\bibitem[Jennings(2008)]{Jen08}
Jennings, P. J. (2008). Using cluster analysis to define geographical rating territories. Applying Multivariate Statistical Models, 34.


\bibitem[Jørgensen \& Paes De Souza(1994)]{Jor94}
Jørgensen, B., \& Paes De Souza, M. C. (1994). Fitting Tweedie's compound Poisson model to insurance claims data. Scandinavian Actuarial Journal, 1994(1), 69-93.


% \bibitem[Kaźmierczak(2007)]{Kaz07}  % uncited – commented out
% Kaźmierczak, A., \& Cavan, G. (2011). Surface water flooding risk to urban communities: Analysis of vulnerability, hazard and exposure. Landscape and urban planning, 103(2), 185-197.

% \bibitem[Krainski et~al.(2018)]{Kra18}  % uncited – commented out
% Krainski, E., Gómez-Rubio, V., Bakka, H., Lenzi, A., Castro-Camilo, D., Simpson, D., ... \& Rue, H. (2018). Advanced spatial modelling with stochastic partial differential equations using R and INLA. Chapman and Hall/CRC.

% \bibitem[Legrand et~al.(2025)Legrand, Naveau, \& Oesting]{Leg25}  % uncited – commented out
% Legrand, J., Naveau, P., \& Oesting, M. (2025). Evaluation of binary classifiers for asymptotically dependent and independent extremes. Journal of the American Statistical Association, 1-19.

\bibitem[Legrand et~al.(2024)Legrand, Pimont, Dupuy, \& Opitz]{Leg24}
Legrand, J., Pimont, F., Dupuy, J. L., \& Opitz, T. (2024). Bayesian spatiotemporal modelling of wildfire occurrences and sizes for projections under climate change. Computo.


\bibitem[Lindgren et~al.(2011)Lindgren, Rue, \& Lindström]{Lin11}
Lindgren, F., Rue, H., \& Lindström, J. (2011). An explicit link between Gaussian fields and Gaussian Markov random fields: the stochastic partial differential equation approach. Journal of the Royal Statistical Society Series B: Statistical Methodology, 73(4), 423-498.

\bibitem[Lyubchich \& Gel(2017)]{Lyu17}
Lyubchich, V., \& Gel, Y. R. (2017). Can we weather proof our insurance?. Environmetrics, 28(2), e2433.

\bibitem[Martínez-Gomariz et~al.(2021)]{Mar21}
Martínez-Gomariz, E., Forero-Ortiz, E., Russo, B., Locatelli, L., Guerrero-Hidalga, M., Yubero, D., \& Castan, S. (2021). A novel expert opinion-based approach to compute estimations of flood damage to property in dense urban environments. Barcelona case study. Journal of Hydrology, 598, 126244.


\bibitem[Merz et~al.(2013)Merz, Kreibich, \& Lall]{Mer13}
Merz, B., Kreibich, H., \& Lall, U. (2013). Multi-variate flood damage assessment: a tree-based data-mining approach. Natural Hazards and Earth System Sciences, 13(1), 53-64.

\bibitem[Merz et~al.(2004)Merz, Kreibich, Thieken, \& Schmidtke]{Mer04}
Merz, B., Kreibich, H., Thieken, A., \& Schmidtke, R. (2004). Estimation uncertainty of direct monetary flood damage to buildings. Natural Hazards and Earth System Sciences, 4(1), 153-163.

\bibitem[Mobini et~al.(2021)Mobini, Nilsson, Persson, Becker, \& Larsson]{Mob21}
Mobini, S., Nilsson, E., Persson, A., Becker, P., \& Larsson, R. (2021). Analysis of pluvial flood damage costs in residential buildings-A case study in Malmö. International Journal of Disaster Risk Reduction, 62, 102407.

\bibitem[Moriah et~al.(2026)Moriah, Vermet, Ailliot, Naveau, \& Legrand]{Mor25}
Moriah M., Vermet F., Ailliot P., Naveau P., \& Legrand J. (2026). Contributions of geolocated weather and building related data for insurance assessment of flood risks. arXiv Preprint 2603.02418

% \bibitem[Nychka et~al.(2015)Nychka, Bandyopadhyay, Hammerling, Lindgren, \& Sain]{Nyc15}  % uncited – commented out
% Nychka, D., Bandyopadhyay, S., Hammerling, D., Lindgren, F., \& Sain, S. (2015). A multiresolution Gaussian process model for the analysis of large spatial datasets. Journal of Computational and Graphical Statistics, 24(2), 579-599.

% \bibitem[Opitz et~al.(2020)Opitz, Bonneu, \& Gabriel]{Opi20}  % uncited – commented out
% Opitz, T., Bonneu, F., \& Gabriel, E. (2020). Point-process based Bayesian modelling of space-time structures of forest fire occurrences in Mediterranean France. Spatial Statistics, 40, 100429.

% \bibitem[Orozco-Acosta et~al.(2021)Orozco-Acosta, Adin, \& Ugarte]{Oro21}  % uncited – commented out
% Orozco-Acosta, E., Adin, A., \& Ugarte, M. D. (2021). Scalable Bayesian modelling for smoothing disease risks in large spatial data sets using INLA. Spatial Statistics, 41, 100496.

\bibitem[Parc National des Cévennes(2020)]{PNC20}
Parc National des Cévennes (2020). Des épisodes Cévenol plus fréquents en Val d'Aigoual. \url{https://www.cevennes-parcnational.fr/fr/actualites/des-episodes-cevenols-plus-frequents}.
\bibitem[PPRI(2023)]{PPRI2023}
Direction régionale et interdépartementale de l'environnement, de l'aménagement et des transports (2023). Plan de Prevention des Risques Inondation. Portail de la Direction régionale et interdépartementale de l'environnement, de l'aménagement et des transports. Retrieved from \url{https://www.drieat.ile-de-france.developpement-durable.gouv.fr/les-plans-de-prevention-des-risques-d-inondation-a4750.html}





\bibitem[Quarteroni(1994)]{Qua94}
Quarteroni, A., \& Valli, A. (1994). Numerical approximation of partial differential equations. Berlin, Heidelberg: Springer Berlin Heidelberg.

% \bibitem[Riebler et~al.(2016)Riebler, Sørbye, Simpson, \& Rue]{Rie16}  % uncited – commented out
% Riebler, A., Sørbye, S. H., Simpson, D., \& Rue, H. (2016). An intuitive Bayesian spatial model for disease mapping that accounts for scaling. Statistical methods in medical research, 25(4), 1145-1165.

\bibitem[Rivas-Lopez et~al.(2025)Rivas-Lopez, Matilla-García, Minguez-Salido, \& Bravo-Ovalle]{Riv25}
Rivas-Lopez, M. V., Matilla-García, M., Minguez-Salido, R., \& Bravo-Ovalle, M. A. (2025). Improving Home Insurance Ratemaking with Geographically Weighted Poisson Regression (GWPR) Model: Assessing Water Damage Risk. Applied Spatial Analysis and Policy, 18(1), 38.

\bibitem[Rivas-Lopez et~al.(2021)Rivas-Lopez, Minguez-Salido, Matilla Garcia, \& Echeverria Rey]{Riv21}
Rivas-Lopez, M. V., Minguez-Salido, R., Matilla Garcia, M., \& Echeverria Rey, A. (2021). Contributions from spatial models to Non-Life insurance pricing: an empirical application to water damage risk. Mathematics, 9(19), 2476.


\bibitem[Rue et~al.(2017)]{Rue17}
Rue, H., Riebler, A., Sørbye, S. H., Illian, J. B., Simpson, D. P., \& Lindgren, F. K. (2017). Bayesian computing with INLA: a review. Annual Review of Statistics and Its Application, 4(1), 395-421.

\bibitem[Rue et~al.(2009)Rue, Martino, \& Chopin]{Rue09}
Rue, H., Martino, S., \& Chopin, N. (2009). Approximate Bayesian inference for latent Gaussian models by using integrated nested Laplace approximations. Journal of the Royal Statistical Society Series B: Statistical Methodology, 71(2), 319-392.

% \bibitem[Sampson et~al.(2014)]{Samps14}  % uncited – commented out
% Sampson, C. C., Fewtrell, T. J., O'Loughlin, F., Pappenberger, F., Bates, P. B., Freer, J. E., \& Cloke, H. L. (2014). The impact of uncertain precipitation data on insurance loss estimates using a flood catastrophe model. Hydrology and Earth System Sciences, 18(6), 2305-2324.

\bibitem[Schaefer(1990)]{Sch90}
Schaefer, J. T. (1990). The critical success index as an indicator of warning skill. Weather and forecasting, 5(4), 570-575.

% \bibitem[Scheel et~al.(2013)]{Sch13}  % uncited – commented out
% Scheel, I., Ferkingstad, E., Frigessi, A., Haug, O., Hinnerichsen, M., \& Meze-Hausken, E. (2013). A Bayesian hierarchical model with spatial variable selection: the effect of weather on insurance claims. Journal of the Royal Statistical Society Series C: Applied Statistics, 62(1), 85-100.


% \bibitem[Shewchuk(1996)]{She96}  % uncited – commented out
% Shewchuk, J. R. (1996, May). Triangle: Engineering a 2D quality mesh generator and Delaunay triangulator. In Workshop on applied computational geometry (pp. 203-222). Berlin, Heidelberg: Springer Berlin Heidelberg.

\bibitem[Smyth \& Jørgensen(2002)]{Smy02}
Smyth, G. K., \& Jørgensen, B. (2002). Fitting Tweedie's compound Poisson model to insurance claims data: dispersion modelling. ASTIN Bulletin: The Journal of the IAA, 32(1), 143-157.

\bibitem[Spekkers et~al.(2014)Spekkers, Kok, Clemens, \& Ten Veldhuis]{Spe14}
Spekkers, M. H., Kok, M., Clemens, F. H. L. R., \& Ten Veldhuis, J. A. E. (2014). Decision-tree analysis of factors influencing rainfall-related building structure and content damage. Natural hazards and earth system sciences, 14(9), 2531-2547.

\bibitem[Spekkers et~al.(2011)Spekkers, ten Veldhuis, Kok, \& Clemens]{Spe11}
Spekkers, M., Marie-claire ten Veldhuis, Kok, M., \& Clemens, F. (2011). Analysis of pluvial flood damage based on data from insurance companies in the Netherlands.

\bibitem[Tansar et~al.(2020)Tansar, Babur, \& Karnchanapaiboon]{Tan20}
Tansar, H., Babur, M., \& Karnchanapaiboon, S. L. (2020). Flood inundation modelling and hazard assessment in Lower Ping River Basin using MIKE FLOOD. Arabian Journal of Geosciences, 13(18), 934.

\bibitem[Taylor(1989)]{Tay89}
Taylor, G. C. (1989). Use of spline functions for premium rating by geographic area. ASTIN Bulletin: The Journal of the IAA, 19(1), 91-122.

\bibitem[Torgersen et~al.(2017)Torgersen, Rød, Kvaal, Bjerkholt \& Lindholm]{Tor17}
Torgersen, G., Rød, J. K., Kvaal, K., Bjerkholt, J. T., \& Lindholm, O. G. (2017). Evaluating flood exposure for properties in urban areas using a multivariate modelling technique. Water, 9(5), 318.

\bibitem[Tufvesson et~al.(2019)Tufvesson, J. Lindström \& Lindström]{Tuf19}
Tufvesson, O., Lindström, J., \& Lindström, E. (2019). Spatial statistical modelling of insurance risk: a spatial epidemiological approach to car insurance. Scandinavian Actuarial Journal, 2019(6), 508-522.

% \bibitem[Van Niekerk et~al.(2021)Van Niekerk, Bakka, Rue, \& Schenk]{Van21}  % uncited – commented out
% Van Niekerk, J., Bakka, H., Rue, H., \& Schenk, O. (2021). New frontiers in Bayesian modelling using the INLA package in R. Journal of Statistical Software, 100, 1-28.

% \bibitem[Verrall(1996)]{Ver96}  % uncited – commented out
% Verrall, R. J. (1996). A unified framework for graduation.

\bibitem[Wahl et~al.(2022)Wahl, Aanes, Aas, Froyn, \& Piacek]{Wah22}
Wahl, J. C., Aanes, F. L., Aas, K., Froyn, S., \& Piacek, D. (2022). Spatial modelling of risk premiums for water damage insurance. Scandinavian Actuarial Journal, 2022(3), 216-233.

\bibitem[Watanabe(2010)]{Wat10}
Watanabe, S., Opper, M. (2010). Asymptotic equivalence of Bayes cross validation and widely applicable information criterion in singular learning theory. Journal of machine learning research, 11(12).


% \bibitem[Woo(2011)]{Woo11}  % uncited – commented out
% Wood, S. N. (2011). Fast stable restricted maximum likelihood and marginal likelihood estimation of semiparametric generalized linear models. Journal of the Royal Statistical Society Series B: Statistical Methodology, 73(1), 3-36.

% \bibitem[Xie(2019)]{Xie19}  % uncited – commented out
% Xie, S. (2019). Defining geographical rating territories in auto insurance regulation by spatially constrained clustering. Risks, 7(2), 42.



\end{thebibliography}
\end{document}